\documentclass[apj]{emulateapj}

\def\ergs{erg~s$^{-1}$}
\def\ergcms{erg~cm$^{-2}$~s$^{-1}$}
\def\fnu{erg~cm$^{-2}$~s$^{-1}$~Hz$^{-1}$}

\begin{document}
\title{Compact Optical Counterparts of Ultraluminous X-ray Sources}

\author{Lian Tao\altaffilmark{1,2}, Hua Feng\altaffilmark{1}, Fabien Gris\'e\altaffilmark{2}, and Philip Kaaret\altaffilmark{2}}
\altaffiltext{1}{Department of Engineering Physics and Center for Astrophysics, Tsinghua University, Beijing 100084, China}
\altaffiltext{2}{Department of Physics and Astronomy, University of Iowa, Van Allen Hall, Iowa City, IA 52242, USA}

\shorttitle{HST Photometry of ULX Counterparts}
\shortauthors{Tao et al.}

\begin{abstract}
Using archival Hubble Space Telescope (HST)  imaging data, we report the multiband photometric properties of 13 ultraluminous X-ray sources (ULXs) that have a unique compact optical counterpart.  Both magnitude and color variation are detected at time scales of days to years.  The optical color, variability, and X-ray to optical flux ratio indicate that the optical emission of most ULXs is dominated by X-ray reprocessing on the disk, similar to that of low mass X-ray binaries. For most sources, the optical spectrum is a power-law, $F_{\nu} \propto \nu^{\alpha}$ with $\alpha$ in the range 1.0 to 2.0 and the optically emitting region has a size on the order of $10^{12}$~cm.  Exceptions are NGC 2403 X-1 and M83 IXO 82, which show optical spectra consistent with direct emission from a standard thin disk, M101 ULX-1 and M81 ULS1, which have X-ray to optical flux ratios more similar to high-mass X-ray binaries, and IC 342 X-1, in which the optical light may be dominated by the companion star.  Inconsistent extinction between the optical counterpart of NGC 5204 X-1 and the nearby optical nebulae suggests that they may be unrelated. 
\end{abstract}


\section{Introduction}

Ultraluminous X-ray sources (ULXs) are nonnuclear X-ray sources with luminosity above the Eddington limit of stellar mass black holes ($L_{\rm X} > 3 \times 10^{39}$~erg~s$^{-1}$) assuming isotropic emission.  ULXs with fast variability or long-term chaotic flux variation may contain accreting black holes in an attached binary system. They could be powered by accretion onto either intermediate mass black holes of $10^2-10^4$~$M_\sun$ at sub-Eddington rate \citep{col99,mak00,kaa01} or normal or slightly massive stellar black holes of $\lesssim 100$~$M_\sun$ at super or near Eddington rate \citep{wat01,beg02}.

In black hole binaries, the X-ray emission probes the deep potential at regions close to the event horizon while the optical emission is important to constrain their masses, interaction with environment, and evolutionary history of the binary.  Optical observations of ULXs have played a key role in unveiling the nature of these sources.  Many bright ULXs are found to be spatially associated with an optical nebula \citep{pak03}.  These nebulae are large with an extent of tens to hundreds pc, and expanding at a velocity in the order of $10^2$~km~s$^{-1}$.  Most of them are powered by strong shocks created by outflows from the binary hitting the surrounding medium, and in a few cases, extraordinary photoionization is required \citep{kaa04,kaa09a}. The photoionized nebulae can be used as bolometers to extreme ultraviolet and soft X-ray photons, and suggest that the ULX luminosities are indeed high. Point-like optical objects as compact counterparts of ULXs have been identified for 10 sources in the literature. The companion star, intrinsic emission from the disk at large radii, and disk irradiation may all contribute to the optical emission and even dominate at different wavelengths. Binary population synthesis indicates that colors and magnitudes of the optical counterpart of ULXs can help distinguish between stellar mass and intermediate mass black holes, and a sample of 6 sources favors the latter interpretation \citep{mad08}. The continuum component in the optical spectrum of NGC 5408 X-1 is consistent with that from a standard accretion disk with irradiation \citep{kaa09a}. However, this result is obtained from a ground-based telescope so the nebular contribution cannot be directly resolved and has to be subtracted based on model prediction. Therefore, in most cases, HST observations are required for the study of the compact optical counterpart of ULXs, in particular for the continuum emission.

There are 10 ULXs with a unique optical counterpart identified in the literature.  We make 3 new identifications in this paper. Here, we analyze all available HST imaging data for these 13 ULXs, see Table~\ref{tab:sample}, to investigate their multiband spectra and variation at long time scales.  The observations and data analysis are described in \S~\ref{sec:obs}, and the results are discussed in \S~\ref{sec:dis} and conclusions summarized in \S~\ref{sec:con}.

\section{Observations and Data Analysis}
\label{sec:obs}

\subsection{Sample}

Up to beginning of analysis work for this paper, there have been 10 ULXs with a unique optical counterpart identified:
Holmberg II X-1 \citep{kaa04},
Holmberg IX X-1 \citep{gri11},
IC 342 X-1 \citep{fen08},
M81 ULS1 \citep{liu08a},
M81 X-6 \citep{liu02},
M101 ULX-1 \citep{kun05},
NGC 1313 X-2 \citep{ram06,pak06,liu07},
NGC 5204 X-1 \citep{liu04},
NGC 5408 X-1 \citep{lan07},
and NGC 6946 ULX-1 \citep{kaa10}.
Moreover, \citet{pta06} and \citet{rob08} reported possible optical counterparts of another 50 ULXs. However, these sources appear to have more than one optical counterparts shown in the X-ray error circle. Here, using Chandra and HST data up to date, we surveyed ULXs in these catalogs for possibilities of improving the relative astrometry, which leads to the identification of a unique counterpart for another 3 sources, M83 IXO 82, NGC 2403 X-1, and NGC 4559 X-7.

%%%%%%%%%%%%%%%%%%%%%%%%%%%%%%%%%%%%%%%%%%%%%%%%%%%%%%%%%%%%%%%%%%%%%%%%%%
\begin{deluxetable}{lllll}
\tablewidth{0pc}
\tabletypesize{\scriptsize}
\tablecaption{Positions of the reference objects, ULXs, and their unique optical counterparts \label{tab:cpt}}
\tablehead{
\colhead{source} & \colhead{instrument} & \colhead{R.A.} & \colhead{decl.} & \colhead{$\Delta$}\\
\colhead{} & \colhead{} & \colhead{(J2000.0)} & \colhead{(J2000.0)} & \colhead{ (\arcsec)}
}
\startdata
\multicolumn{5}{c}{NGC 4559 X-7}\\
\hline\noalign{\smallskip}
Ref.\ 1 & Chandra & 12 36 00.353 & +27 54 56.61 &  0.271 \\
 & HST & 12 36 00.354 & +27 54 56.67 & \\
ULX & Chandra & 12 35 51.724 & +27 56 04.35 &  0.052 \\
& corrected & 12 35 51.725  & +27 56 04.41 & 0.28 \\
counterpart & HST  & 12 35 51.719 & +27 56 04.41  & \\
\hline\noalign{\smallskip}
\multicolumn{5}{c}{M83 IXO 82}\\
\hline\noalign{\smallskip}
Ref.\ 1& Chandra & 13 37 14.764 & $-$29 54 28.47 & 0.318 \\
& HST & 13 37 14.761 & $-$29 54 28.26 & \\
Ref.\ 2& Chandra & 13 37 19.986 & $-$29 53 18.09 & 0.169 \\
& HST & 13 37 19.979 & $-$29 53 17.68 & \\
ULX & Chandra & 13 37 19.804 & $-$29 53 48.80 & 0.029 \\
& corrected & 13 37 19.799 & $-$29 53 48.43 & 0.18 \\
counterpart & HST & 13 37 19.797 & $-$29 53 48.58 & \\
\hline\noalign{\smallskip}
\multicolumn{5}{c}{NGC 2403 X-1}\\
\hline\noalign{\smallskip}
ULX & Chandra & 07 36 25.567 &  +65 35 39.87  &  0.6 \\
 & Chandra & 07 36 25.558  &  +65 35 39.98 &  0.6 \\
& mean &  07 36 25.563 & +65 35 39.93  &  0.42 \\
counterpart & HST &  07 36 25.562 & +65 35 40.08  &
\enddata
\tablecomments{The used Chandra observation is 2686 for NGC 4559, 793 for M83, and 2014/4630 for NGC 2403. The used HST datasets are j92w02030, j9h811010, and j9h803010, respectively for the three targets. $\Delta$ is the 90\% statistical error radius; for NGC 2403 X-1, it is the absolute astrometry error of Chandra. }
\end{deluxetable}
%%%%%%%%%%%%%%%%%%%%%%%%%%%%%%%%%%%%%%%%%%%%%%%%%%%%%%%%%%%%%%%%%%%%%%%%%%

%%%%%%%%%%%%%%%%%%%%%%%%%%%%%%%%%%%%%%%%%%%%%%%%%%%%%%%%%%%%%%%%%%%%%%%%%%
\begin{figure*}
\centerline{\includegraphics[width=2.1\columnwidth]{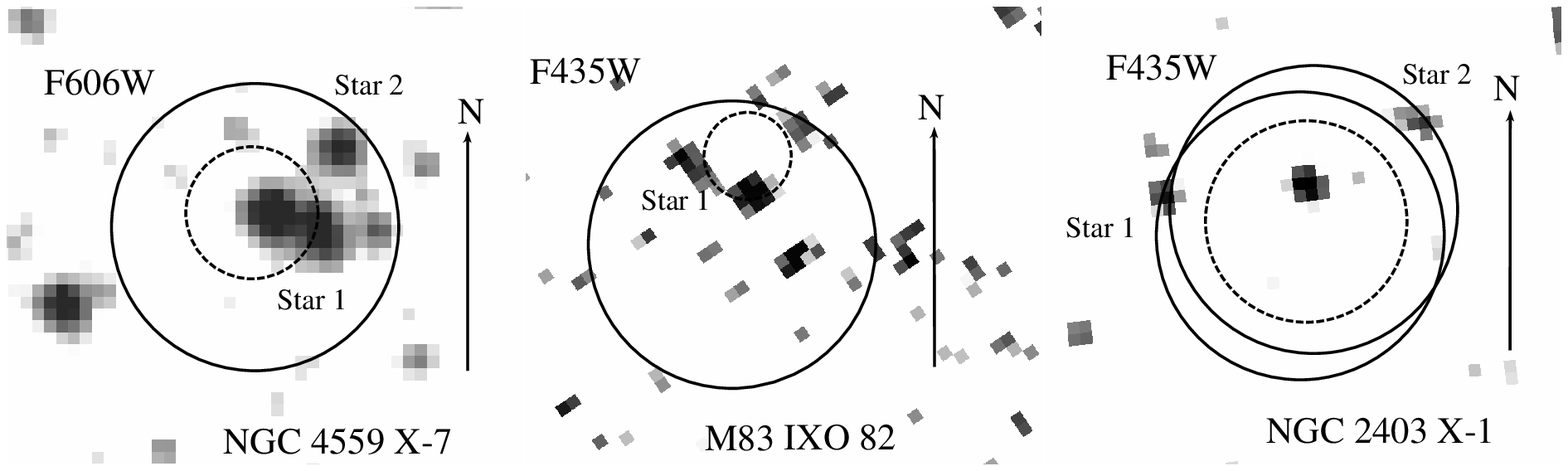}\\}
\includegraphics[width=0.7\columnwidth]{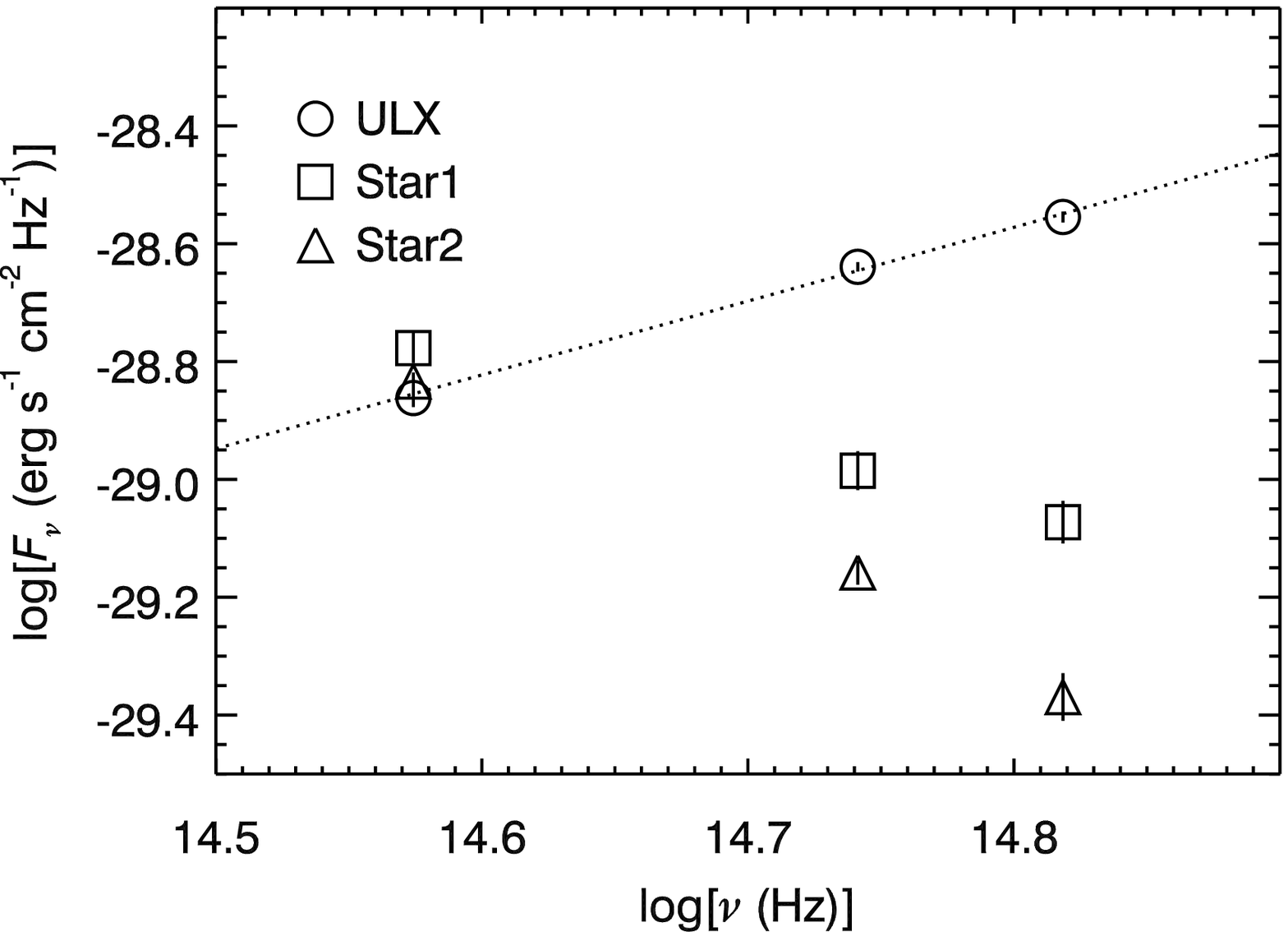} 
\includegraphics[width=0.7\columnwidth]{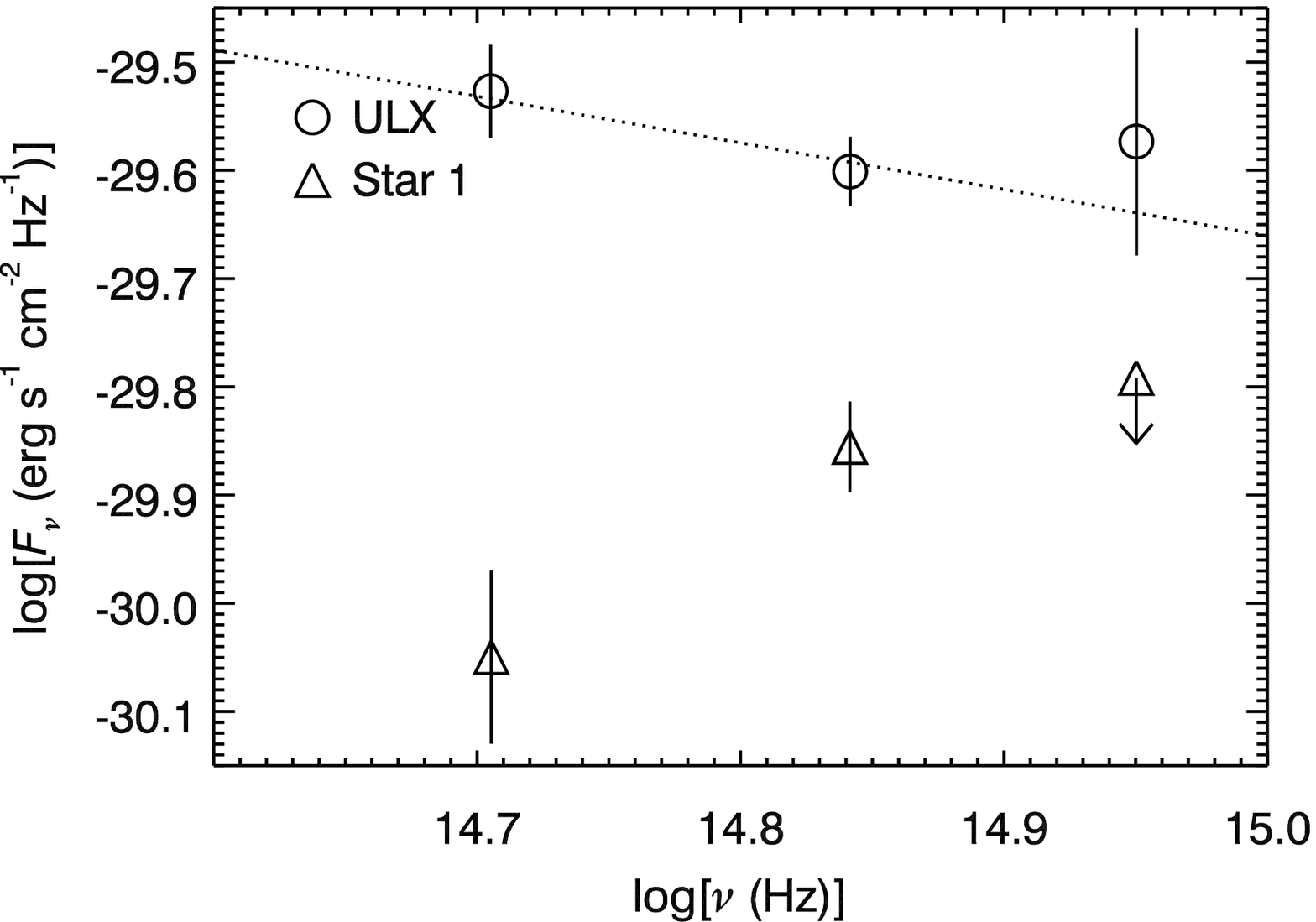}
\includegraphics[width=0.7\columnwidth]{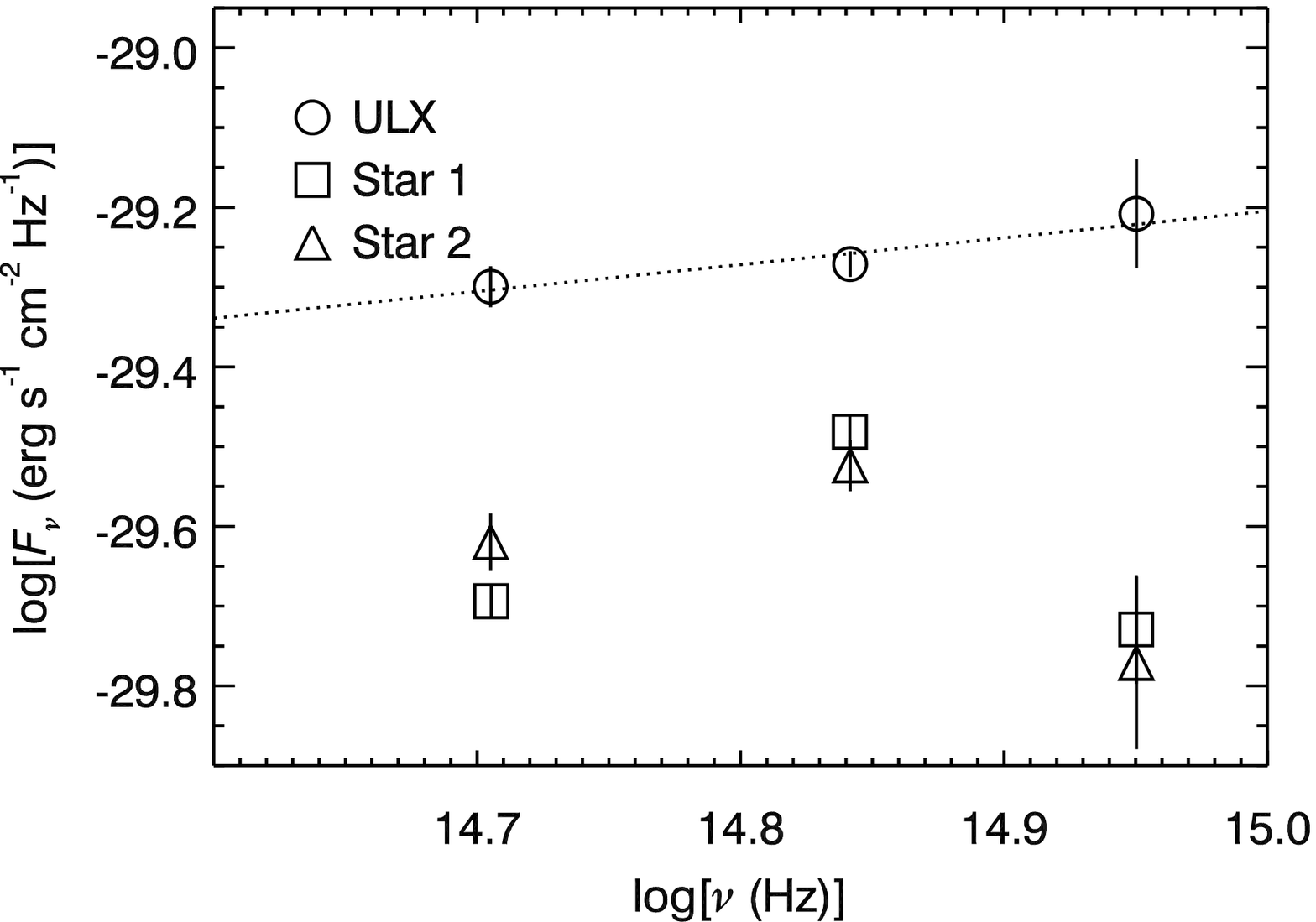}

\caption{{\bf Upper panels:} HST images around NGC 4559 X-7, M83 IXO 82 and NGC 2403 X-1 . The solid circle indicates the original X-ray position of the ULX, with a radius of 0.6\arcsec\ corresponding to the absolute astrometry error of Chandra. The arrow points north and has a length of 1\arcsec.  For NGC 4559 X-7 and M83 IXO 82, the dashed circle is the corrected X-ray position after alignment. For NGC 2403 X-1, the dashed circle is the mean position of the two Chandra observations. A unique optical counterpart is identified for each source.  {\bf Lower panels:} SEDs of the ULXs and nearby sources.  The nearby sources are marked on the images in the upper panel.  In all three cases, the SED of the ULX counterpart is consistent with a single power-law, while the nearby stars show a turn over at short wavelengths. }
\label{fig:align}
\end{figure*}
%%%%%%%%%%%%%%%%%%%%%%%%%%%%%%%%%%%%%%%%%%%%%%%%%%%%%%%%%%%%%%%%%%%%%%%%%%

The details about the astrometry correction have been described in \citet{fen08}. The relative astrometry between Chandra and HST is improved by aligning objects (hereafter referred as reference objects) shown on both. For Chandra, deepest observations with the Advanced CCD Imaging Spectrometer are used. The X-ray positions are calculated using the {\tt wavdetect} tool in CIAO from exposure-corrected images in the energy range of 0.3--8 keV.  The HST images are first registered to the Two-Micron All Sky Survey \citep[2MASS;][]{skr06} source frame to achieve better absolute astrometry before searching for reference objects. For NGC 4559 X-7 and M83 IXO 82, there is 1 and 2 reference objects, respectively, identified for image alignment. The corrected X-ray error circle of the ULX after alignment, taking into account of all uncertainties, is around 0.2\arcsec-0.3\arcsec, leading to a unique identification (Table~\ref{tab:cpt}). For NGC 2403 X-1, although we cannot find a reference object to align the images, there is only one HST object shown around the mean position from two Chandra observations. We thus regard this object as the optical counterpart of the X-ray source.  The HST images around the 3 ULXs are shown in Figure~\ref{fig:align}. The solid circle indicates the 90\% absolute position error of the ULX from Chandra with a radius of 0.6\arcsec, and the dashed circle indicates the aspect-corrected X-ray position relative to optical images with a radius listed in Table~\ref{tab:cpt}. The uncertainty in the ULX error circle is calculated via error propagation using the uncertainties in the position of the ULX relative to the reference source(s) in both the X-ray and optical.  This alignment technique, based on a small number of reference sources, has been used for most ULX optical identifications.  In several cases, the optical counterpart identification has been confirmed via the detection of high excitation emission lines that could not be produced by a star with the given optical colors, but are easily produced by X-ray illumination of an accretion disk.

%  To investigate further the possibility of misidentification, we examined the spectral energy distributions of other stars near the ULX error circles as shown in the lower panels of Fig.~\ref{fig:align}.  In all three cases, the SED of the ULX counterpart is consistent with a single power-law, while the nearby stars show a turn over at short wavelengths.  As discussed further below, this is a common property of ULX optical counterparts and lends credence to our counterpart identifications.}

As a whole, a list of the 13 ULXs in our sample with related information, like the distance to their host galaxies, Galactic extinction along the line of sight, their environmental associations, and the references, is shown in Table~\ref{tab:sample}.

%%%%%%%%%%%%%%%%%%%%%%%%%%%%%%%%%%%%%%%%%%%%%%%%%%%%%%%%%%%%%%%%%%%%%%%%%%
\begin{deluxetable}{llllcl}
\tablewidth{\columnwidth}
\tabletypesize{\scriptsize}
\tablewidth{0pc}
\tablecolumns{6}
\scriptsize
\tablecaption{List of the ULXs studied in this paper.
\label{tab:sample}}
\tablehead{
\colhead{ULX} & \colhead{$D$} & \colhead{$E_{\rm g}$} & \colhead{$E_{\rm n}$} & \colhead{Neb.} & \colhead{Ref.} \\
\colhead{(1)} & \colhead{(2)} & \colhead{(3)} & \colhead{(4)} & \colhead{(5)} & \colhead{(6)}}

\startdata
Holmberg II X-1                & 3.05 & 0.032           & 0.07            & Y   & 1-4\\
Holmberg IX X-1                & 3.6  & 0.079           & $0.26\pm0.04$   & Y   & 1,2,5,6\\
IC 342 X-1                     & 3.3  & 0.565           & 0.82            & Y   & 1,2,7,8\\
M81 ULS1                       & 3.63 & 0.080           & \nodata         & N   & 1,5\\
M81 X-6                        & 3.63 & 0.080           & \nodata         & Y   & 1,2,5 \\
M83  IXO 82                    & 4.7  & 0.066           & \nodata         & N   & 1,9\\
M101 ULX-1                     & 7.2  & 0.009           &0.13             & Y   & 1,10,11\\
NGC 1313 X-2                   & 3.7  & 0.085           & $0.13\pm0.03$   & Y   & 1,2,9,12\\
NGC 2403 X-1                   & 3.2  & 0.040           & \nodata         & N   & 1,13\\
NGC 4559 X-7                   & 10   & 0.018           & \nodata         & N   & 1,9\\
NGC 5204 X-1                   & 4.3  & 0.013           & \nodata         & Y   & 1,2,14 \\
NGC 5408 X-1                   & 4.8  & 0.069           & $0.08\pm0.03$   & Y   & 1,15-17\\
NGC 6946 ULX-1                 & 5.1  & 0.343           & $0.50^{+0.08}_{-0.07}$   & Y   & 1,18-22
\enddata
\tablecomments{$D$ is the adopted distance to the host galaxy in Mpc. $E_{\rm g}$ is the Galactic extinction and $E_{\rm n}$ is the total extinction estimated from the surrounding optical nebulae. Column (5) indicates whether or not the source is spatially associated with an optical nebula. Column (6) lists the references from where the information is obtained:
$^1$\citealt{sch98}; 
$^2$\citealt{pak02}; 
$^3$\citealt{ste00}; 
$^4$\citealt{hoe98}; 
$^5$\citealt{fre94}; 
$^6$\citealt{gri11}; 
$^7$\citealt{gri06}; 
$^8$\citealt{sah02}; 
$^9$\citealt{tul88}; 
$^{10}$\citealt{abo07}; 
$^{11}$\citealt{ste98};
$^{12}$\citealt{gri08};
$^{13}$\citealt{fre88};
$^{14}$\citealt{tul92};
$^{15}$\citealt{kar02};
$^{16}$\citealt{kaa09a};
$^{17}$\citealt{pak03};
$^{18}$\citealt{dev79};
$^{19}$\citealt{abo08};
$^{20}$\citealt{bla01};
$^{21}$\citealt{rob03};
$^{22}$\citealt{bla94}.}
\end{deluxetable}
%%%%%%%%%%%%%%%%%%%%%%%%%%%%%%%%%%%%%%%%%%%%%%%%%%%%%%%%%%%%%%%%%%%%%%%%%%

\subsection{Photometry}

% aperture & background
The HST data used in the paper are observations with either the Advanced Camera for Surveys (ACS) instrument or the Wide-Field Planetary Camera 2 (WFPC2). For ACS data, the photometry is performed on drizzled images produced from standard pipeline calibration using the APPHOT package in IRAF with an aperture radius of 0.15\arcsec-0.2\arcsec.  For Holmberg IX X-1, IC 342 X-1, and NGC 4559 X-7, each has a weak optical source located at about 0.2\arcsec\ from the counterpart of the ULX. To minimize the contamination, we used a smaller aperture with radius of 0.1\arcsec-0.15\arcsec\ for photometry and added the aperture correction error into the results.  For WFPC2 data, the photometry is performed on the c0f images using HSTPHOT 1.1, which outputs aperture-corrected VEGA magnitudes.  As some sources are shown in crowded regions, the choice of background may slightly alter the derived net flux, which is particularly prominent for sources associated with an optical nebula that a flux difference up to 30\% could arise whether or not the nebular emission is subtracted \citep[e.g.][]{kaa10}. Therefore, the uncertainty introduced by sky background subtraction is taken into account, estimated using various background regions or background estimate means.

% extinction
The total extinction along the line of sight to the source is adopted in two ways. For those associated with an optical nebula and the extinction to the nebula has been measured from emission lines, the nebular extinction is adopted (see Table~\ref{tab:sample}). Otherwise, we adopt Galactic extinction \citep{sch98} for reddening correction.  The extinction can also be estimated from the neutral hydrogen column density, which is derived via X-ray spectral fitting.  For Holmberg II X-1, IC 342 X-1, M101 ULX-1, NGC 1313 X-2, NGC 5408 X-1,  and NGC 6946 ULX-1, which reside in nebulae, the ratio of Galactic to nebular extinctions ($E_{\rm g}/E_{\rm n}$) are 0.46, 0.69, 0.07, 0.65, 0.86, and 0.69, respectively, and the ratio of X-ray to nebular extinctions ($E_{\rm x}/E_{\rm n}$) are 1.0, 1.42, 0.85, 3.0, 1.63, and 1.42, respectively.  For Holmberg IX X-1, the nebular extinction is found to vary with position within the nebula from 0.09 to 0.26 \citep{gri11}, and consequently causing $E_{\rm g}/E_{\rm n}$ to vary between 0.3 and 0.9, and $E_{\rm x}/E_{\rm n}$ between 1.3 and 3.9. It seems that $E_{\rm x}$ often overestimate the total extinction. Plus, the X-ray column density usually varies dramatically, causing additional difficulty in deriving the extinction. Thus, we adopt Galactic extinction when the nebular extinction is unavailable, and this should be treated as the lower limit. A Galactic extinction law \citep{car89} is assumed for reddening correction.  Except for the small handful of images in the near UV (F330W filter), the resultant magnitude varies very little, by 0.002 typically, if we choose a different reddening law for the excessive extinction above the Milky Way, like that for the Small Magellanic Clouds where the metallicity is significantly lower.  We hence use the Galactic reddening law for all objects despite that metal poor environments have been suggested for a few of them \citep[e.g.][]{mir02,sor05}.

% spectral shape, & fitting, discussion of adjacent stars
A power-law spectrum is assumed to convert from the observed count rate or magnitude to the intrinsic flux or magnitude.  We fit the intrinsic multiband spectrum after reddening correction with a power-law function, $F_{\nu} \propto \nu^{\alpha}$, where $\nu$ is the pivot frequency of the filter in Hz and $F_{\nu}$ is the flux in erg~cm$^{-2}$~s$^{-1}$~Hz$^{-1}$. The same $\alpha$ is, in return, used for the assumed spectrum. A couple times of iterations are needed in the calculation until the fitted $\alpha$ is consistent with the input $\alpha$. Because we are only interested in the continuum component of the optical spectrum, data from narrow and medium bands are not used due to the possibility that the flux is dominated by emission lines. The optical spectrum of ULXs may be highly variable, thus only quasi-simultaneous observations executed within 2 days are used in the fits except for Holmberg II X-1, M81 ULS1, and NGC 5204 X-1, whose quasi-simultaneous observations only cover two bands, and spectra from different epochs are consistent with a single power-law. A power-law relation is adequate to fit all spectra in the sample, except for a few sources where excessive flux in the F814W band is seen. We compared the properties of the ULX counterparts with nearby stars for the three newly identified sources, NGC 4559 X-7, M83 IXO 82 and NGC 2403 X-1, in Figure~\ref{fig:align}.  The ULX counterparts differ from the nearby stars in that their spectral energy distributions (SEDs) are well fitted with a single power-law, while those of the nearby stars show a roll-over at short wavelengths suggesting a blackbody-like spectrum with a moderate temperature.  Also, the nearby stars are less luminous than the ULX counterpart.  These trends are common to the whole sample. The dereddened magnitude in the observing band for the compact optical counterpart of each ULX from all observations are listed in the Appendix (Table~\ref{tab:mag}).

\subsection{Spectral Classification}

Table~\ref{tab:sp} gives the absolute visual magnitudes and color indices for each quasi-simultaneous observation with at least three filters for each source, except for Holmberg II X-1, M81 ULS1, and NGC 5204 X-1, for which only two bands are available quasi-simultaneously and all their observations are listed.  The most likely stellar spectral classification was inferred for each observation based on the magnitude and colors.  For most sources, the classifications are not consistent with all the colors within one observation or the data from different observations suggest different spectral types for a given source.  The exceptions are IC 342 X-1 and NGC 4559 X-7, which appear to be consistent with a fixed spectral type.

As a further check, we compared stellar models for the candidate classifications to all of the available data for each source and calculated the $\chi^2$ deviation of the (non-simultaneous) data from the stellar model normalized to the average V-magnitude.  For NGC 1313 X-2 and M101 ULX-1, where there are a large number of observations available, we chose a subset of the data.  The $\chi^2$ and degrees of freedom (dof) are reported in the table.  In most cases, the fit to the stellar model is strongly excluded by the data.  The exceptions are IC 342 X-1 and NGC 6946 X-1.  IC 342 X-1 appears consistent with an F5I star both from the stellar spectral fit and the colors noted above.  NGC 6946 X-1 has a low $\chi^2$ because of the large uncertainty in it reddening correction.  The $\chi^2$ of NGC 4559 X-7, which has a consistent spectral classification based on colors, is dominated by variability.

\subsection{Distribution of Spectral Parameters}

\begin{figure}
\includegraphics[width=\columnwidth]{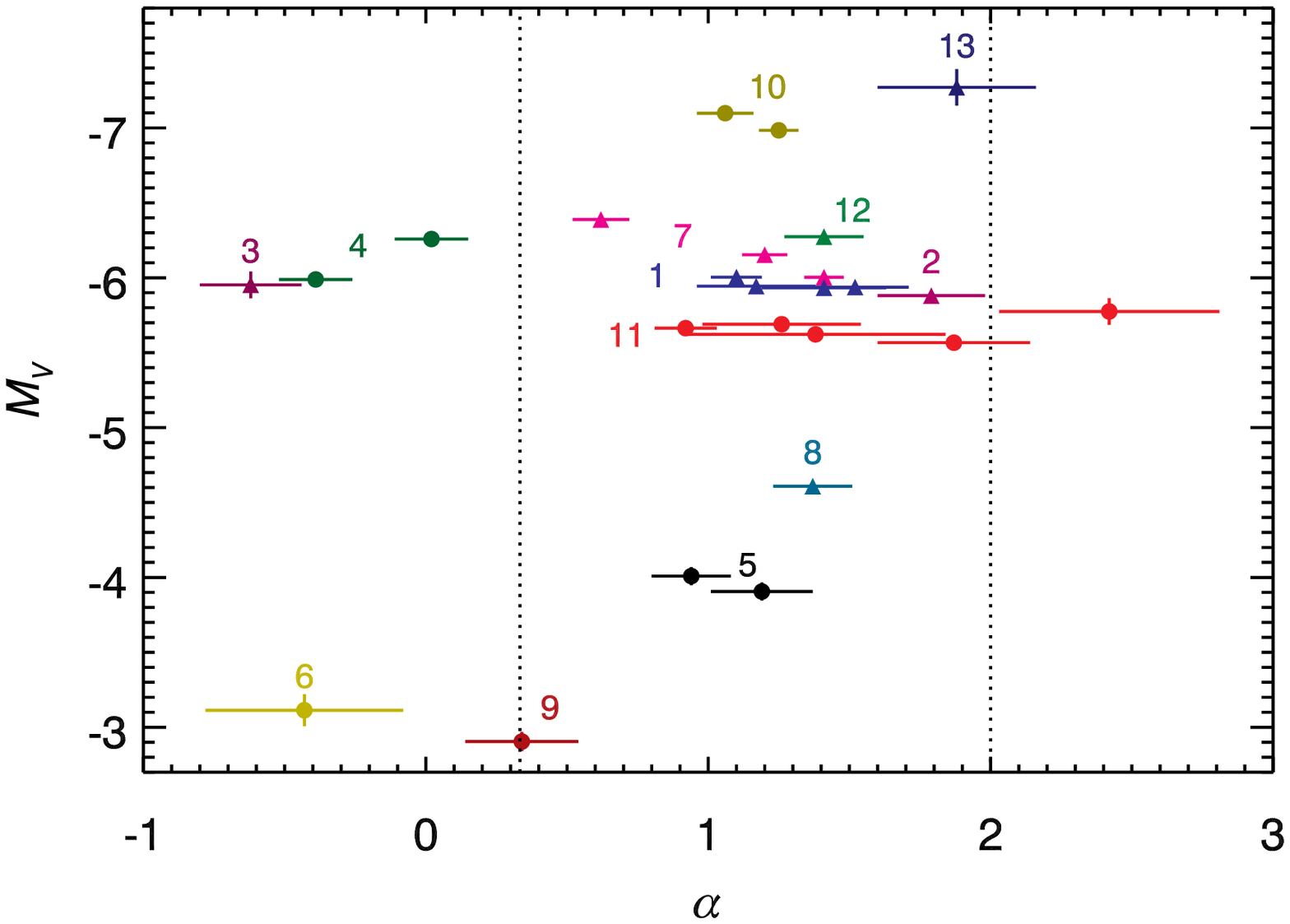} \\
\includegraphics[width=\columnwidth]{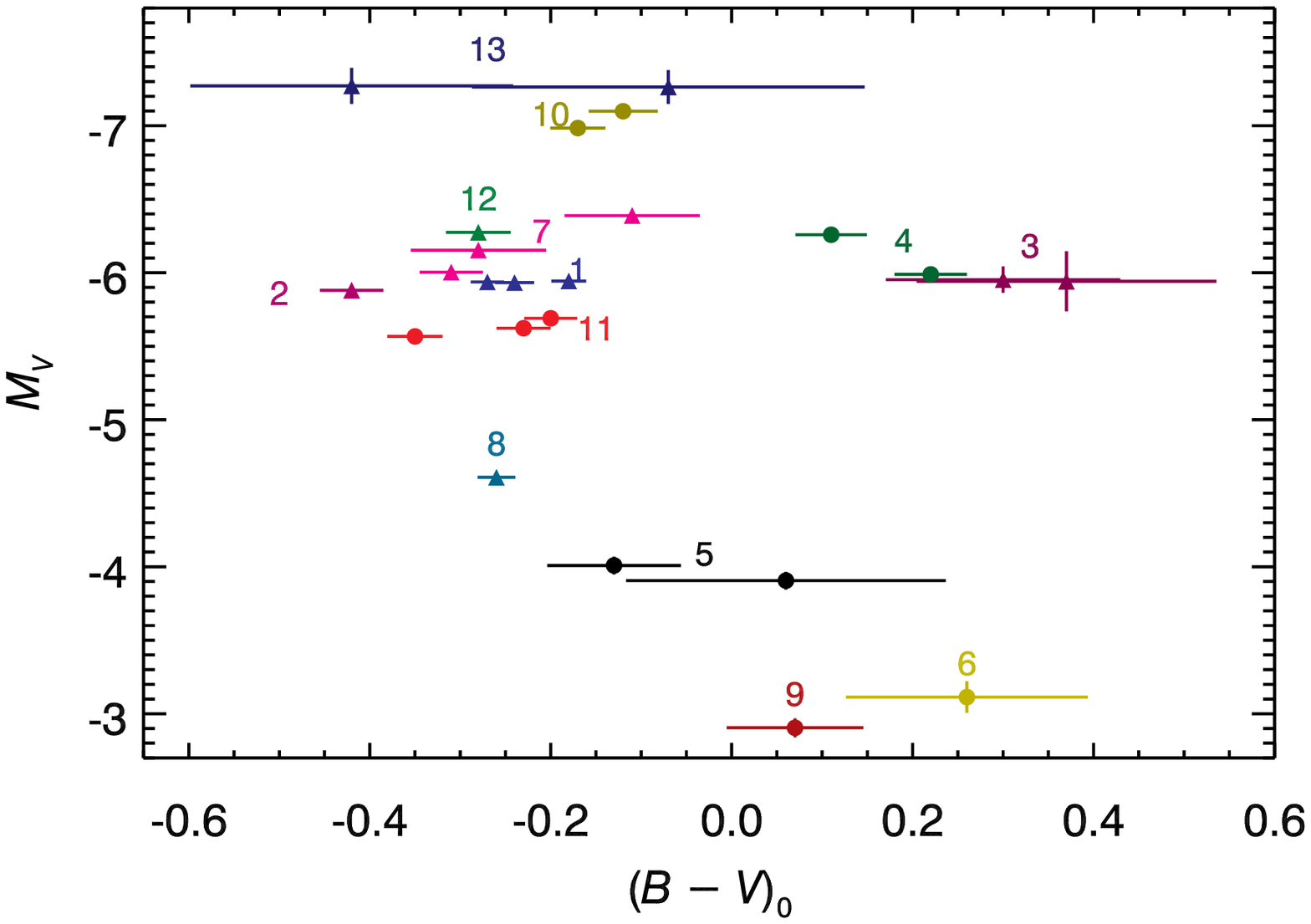}
\caption{The absolute visual magnitude $M_V$ versus the spectral power-law index $\alpha$ (top) or versus $(B-V)_0$ (bottom).  The dotted lines at $\alpha = 1/3$ and $\alpha = 2$ are respectively the values expected for the intrinsic emission from a steady thin disk and for the Rayleigh-Jeans tail of a blackbody. Different points with the same color are from different observations of the same source. Sources for which the reddening was derived from an associated nebula are shown as triangles, while circle denote sources for which only Galactic reddening was assumed.
The sources are:
$^1$Holmberg II X-1;
$^2$Holmberg IX X-1;
$^3$IC 342 X-1;
$^4$M81 ULS1;
$^5$M81 X-6;
$^6$M83 IXO 82;
$^7$M101 ULX-1;
$^8$NGC 1313 X-2;
$^9$NGC 2403 X-1;
$^{10}$NGC 4559 X-7;
$^{11}$NGC 5204 X-1;
$^{12}$NGC 5408 X-1;
$^{13}$NGC 6946 ULX-1.
\label{fig:alpha}}
\end{figure}

\begin{figure}
\includegraphics[width=\columnwidth]{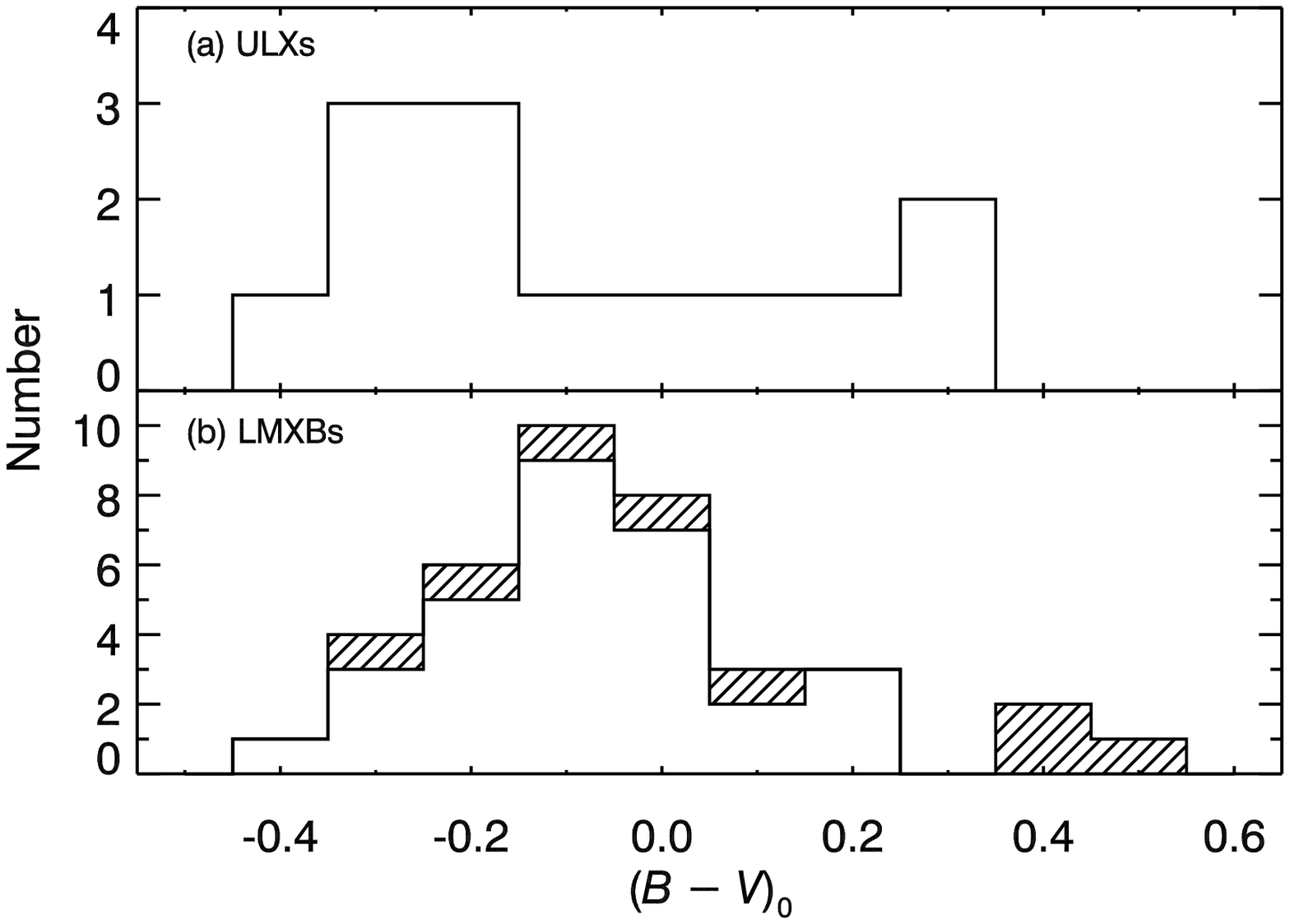} \\
\includegraphics[width=\columnwidth]{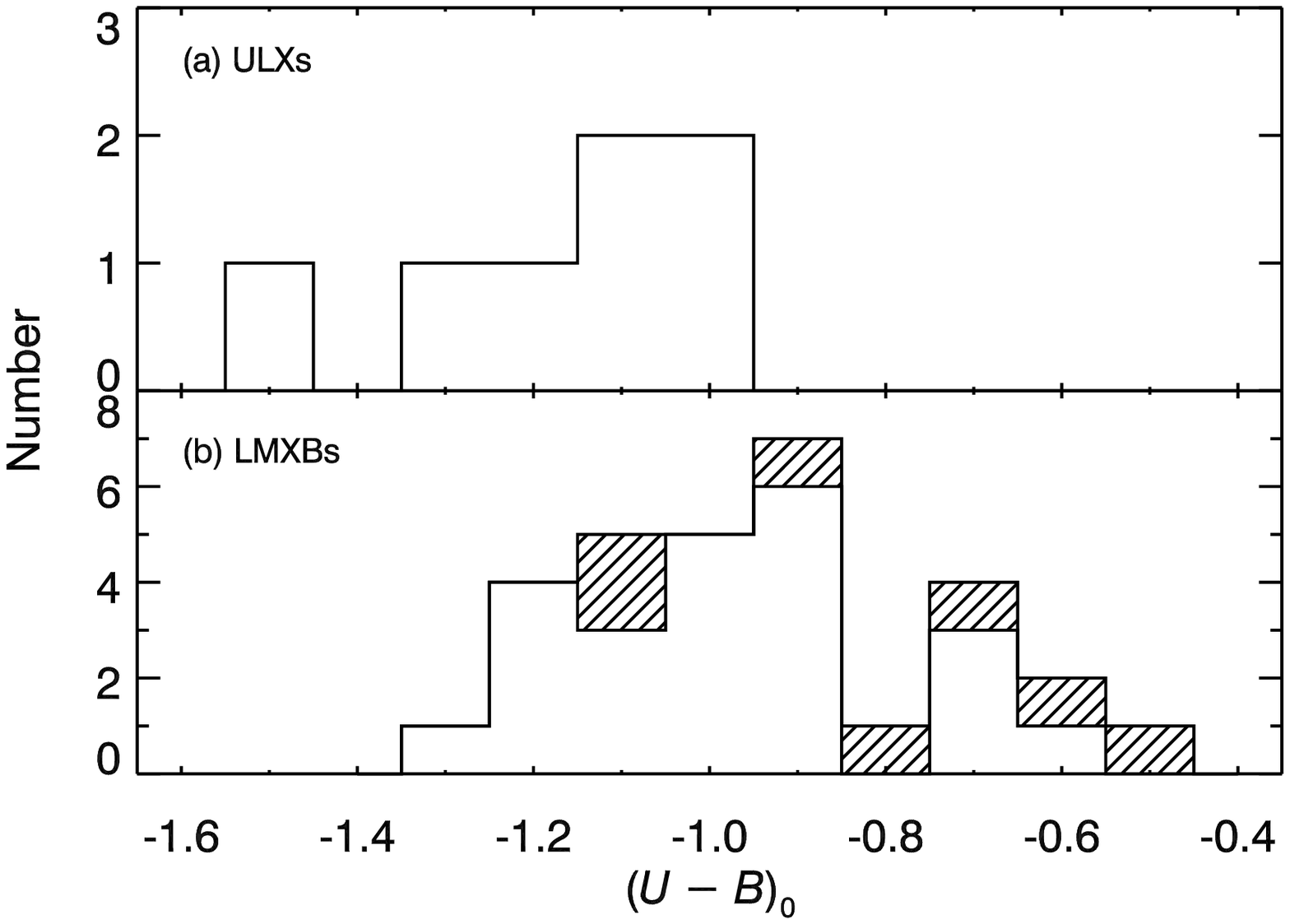}
\caption{The two upper panels are histograms of the average color
indices $\bv$ and $\ub$. The two lower panels are distributions of
$\bv$ and $\ub$ for optical counterparts of LMXBs from
\citet{van95}. The sources indicated in white include some strange
sources in \citet{van95}. \label{fig:color_histogram}}
\end{figure}

% M_V versus alpha and (B-V)_0
Diagrams of $M_V$ versus $\alpha$ and $M_V$ versus $(B-V)_0$ are shown in Figure~\ref{fig:alpha} and the distribution of the optical colors is shown in Fig.~\ref{fig:color_histogram}.  A value for $\alpha$ is computed only when the spectrum has at least three bands, except for Holmberg II X-1, M81 ULS1, and NGC 5204 X-1 whose quasi-simultaneous spectra have only two points.  The absolute magnitude $M_V$ is found in a wide range, from $-7$ to $-3$, for all sources. The spectral index $\alpha$ also appears in a wide range from $-0.6$ to around 2.0, but seems to have a distribution peaked between 1 and 2. The $\alpha$ of NGC 6946 ULX-1, NGC 5204 X-1 and Holmberg IX X-1 is consistent with 2.0, which is the slope of the Rayleigh-Jeans tail of a blackbody spectrum. The $\alpha$ of most sources is less than 2.0, consistent with the fact that no thermal emission can produce a spectrum with $\alpha > 2$. It is worth noting that the derived $\alpha$ is sensitive to the extinction that is assumed; higher extinction leads to larger $\alpha$.  Here, many sources are assumed to have Galactic extinction, which may be lower than the total extinction by 30\% or so (see the $E_{\rm g}/E_{\rm n}$ ratios above). Taking into account of this factor, it is likely that the peak of $\alpha$ distribution is slightly larger and close to 2.  The $\alpha$ expected for the intrinsic emission from a steady thin disk is 1/3, which, however, is inconsistent with most sources except NGC 2403 X-1.

%%%%%%%%%%%%%%%%%%%%%%%%%%%%%%%%%%%%%%%%%%%%%%%%%%%%%%%%%%%%%%%%%%%%%%%%%%
%%%%%%%%%%%%%%%%%%%%%%%%%%%%%%%%%%%%%%%%%%%%%%%%%%%%%%%%%%%%%%%%%%%%%%%%%%

%%%%%%%%%%%%%%%%%%%%%%%%%%%%%%%%%%%%%%%%%%%%%%%%%%%%%%%%%%%%%%%%%%%%%%%%%%
%%%%%%%%%%%%%%%%%%%%%%%%%%%%%%%%%%%%%%%%%%%%%%%%%%%%%%%%%%%%%%%%%%%%%%%%%%

\subsection{Long-term Variation}

% V band variation
F555W and F606W are the two most often used filters in our sample. Magnitudes in Johnson $V$ were calculated from these two bands to examine variability (see Table~\ref{tab:mag}). The observed range in the V-band, $V_{\max} - V_{\min}$, was calculated for each source and is shown in Table~\ref{tab:sum}. 

\begin{figure}
\includegraphics[width=\columnwidth]{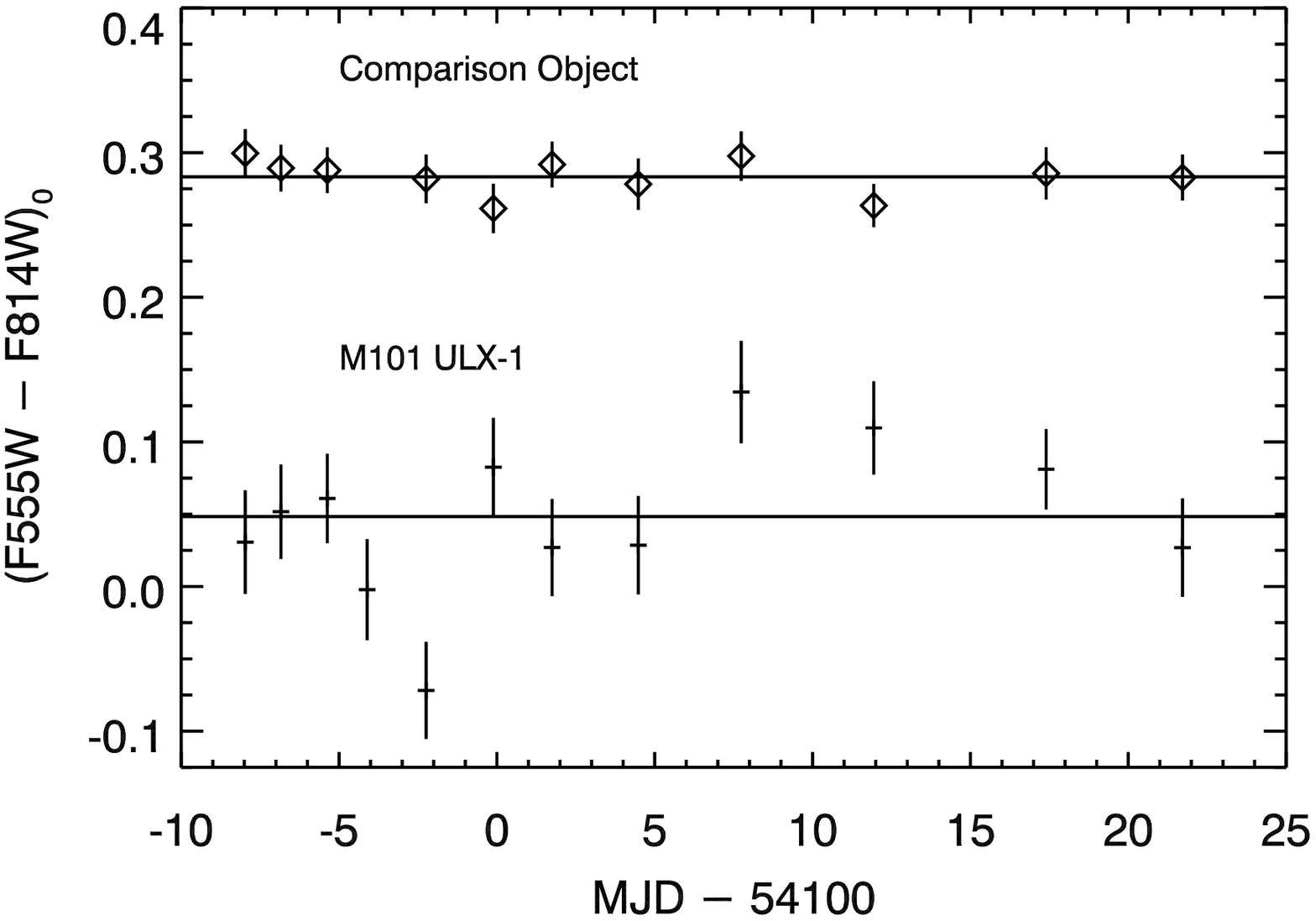}\\
\includegraphics[width=\columnwidth]{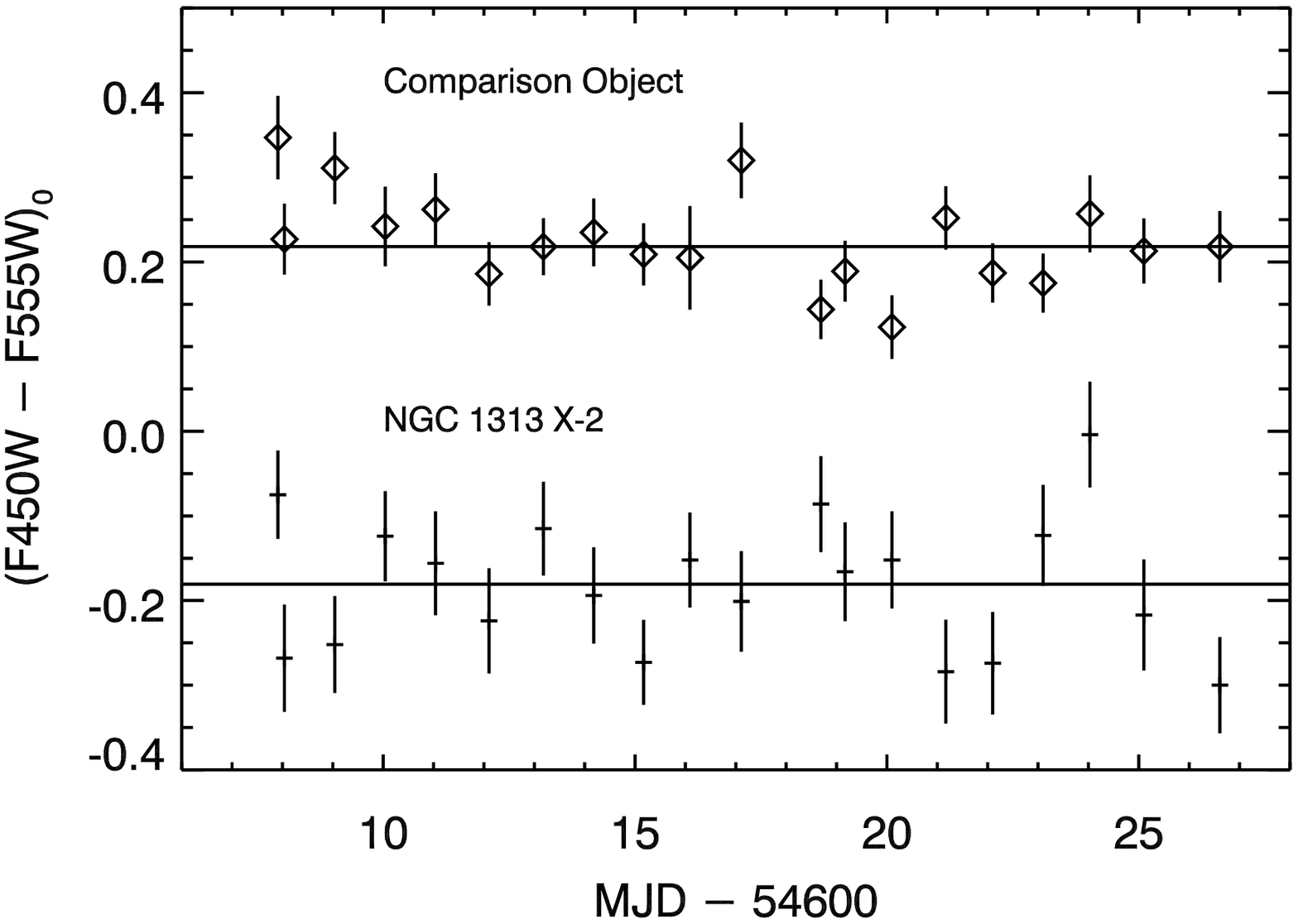}
\caption{Color variation of M101 ULX-1 and NGC 1313 X-2 (plus) and comparison objects (diamond). For the comparison object of M101 ULX-1, its flux on MJD 54096 was affected by cosmic rays and thus not plotted.
\label{fig:var}}
\end{figure}

% M101 and NGC 1313 X-2
For M101 ULX-1 and NGC 1313 X-2, multiple quasi-simultaneous observations in two bands are available, allowing us to investigate long-term variability.  Not only the magnitude, but also the color index varies significantly, see lightcurves of $({\rm F555W} - {\rm F814W})_0$ for M101 ULX-1 and of $({\rm F450W} - {\rm F555W})_0$ for NGC 1313 X-2 in Figure~\ref{fig:var}.  For M101, we plotted ACS data from 2006 Dec 23 to 2007 Jan 21. For NGC 1313 X-2, all WFPC2 data are plotted. We have checked lightcurves obtained using different sky background subtractions and they show the same variation pattern, suggesting that the variability does not arise from the background.  For each ULX, a nearby comparison object was chosen to examine instrumental effects.  For M101 ULX-1, fitting with a constant results in $\chi^2$ per degree of freedom (D.o.F.) of 2.6 and 0.6, respectively for the ULX and the comparison object. For NGC 1313 X-2, the color seems to be constant during the first three quarters of the total observing window, but exhibits large variation in the last 6 observations. Fitting with a constant for the last 6 colors, the $\chi^2 / {\rm d.o.f.}$ is 3.6 and 0.8, respectively for the ULX and the comparison object.

% rule out solely extinction variation
Change of the intrinsic emission or the extinction along the line of sight could both give rise to the observed color variation. Assuming constant intrinsic emission, the change of extinction can be directly derived from the change of observed magnitude in each band. However, for each source, the inferred extinctions at the two bands do not agree with each other. Therefore, pure extinction variation can be ruled out as the nature of the observed variability, and variation of the intrinsic optical emission is required. The irregular nature of the variability indicates that it cannot be completely caused by a periodic phenomena, such as ellipsoidal variations.  Changes in X-ray illumination, perhaps due to changes in the mass accretion rate, could cause the observed variability.  Indeed, for NGC 1313 X-2, as previously shown by \citet{imp10}, the variability is higher in the B-band (F450W) than in the V-band (F555W), suggesting that the variability arises from X-ray irradiation.  Unfortunately, there were no suitable, contemporaneous X-ray observations that would allow us to study the correlation between X-ray and optical variability.

%%%%%%%%%%%%%%%%%%%%%%%%%%%%%%%%%%%%%%%%%%%%%%%%%%%%%%%%%%%%%%%%%%%%%%%%%%
%%%%%%%%%%%%%%%%%%%%%%%%%%%%%%%%%%%%%%%%%%%%%%%%%%%%%%%%%%%%%%%%%%%%%%%%%%

\subsection {X-ray to Optical Flux Ratio}

%%%%%%%%%%%%%%%%%%%%%%%%%%%%%%%%%%%%%%%%%%%%%%%%%%%%%%%%%%%%%%%%%%%%%%%%%%
\begin{figure}
\includegraphics[width=\columnwidth]{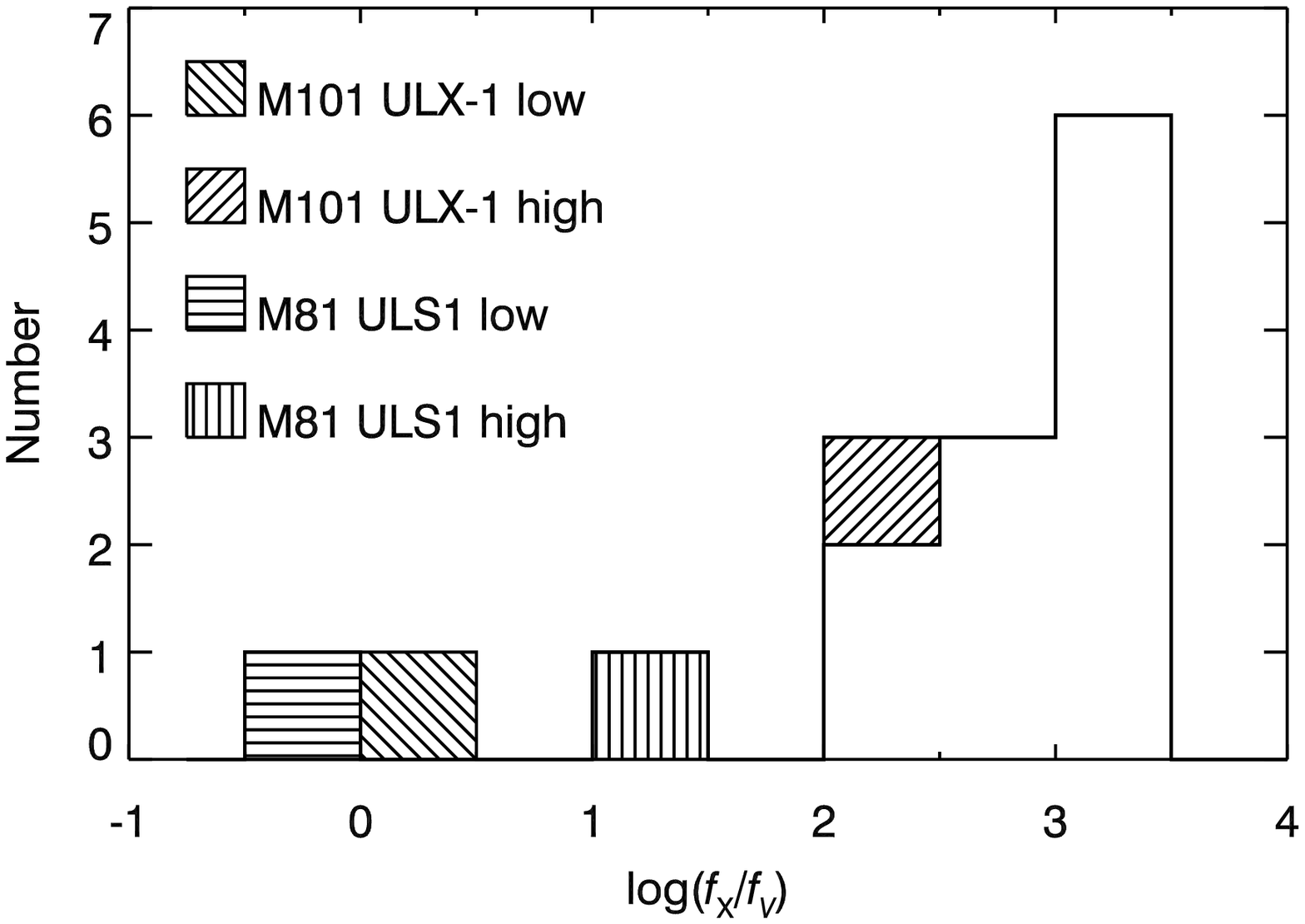}\\
\includegraphics[width=\columnwidth]{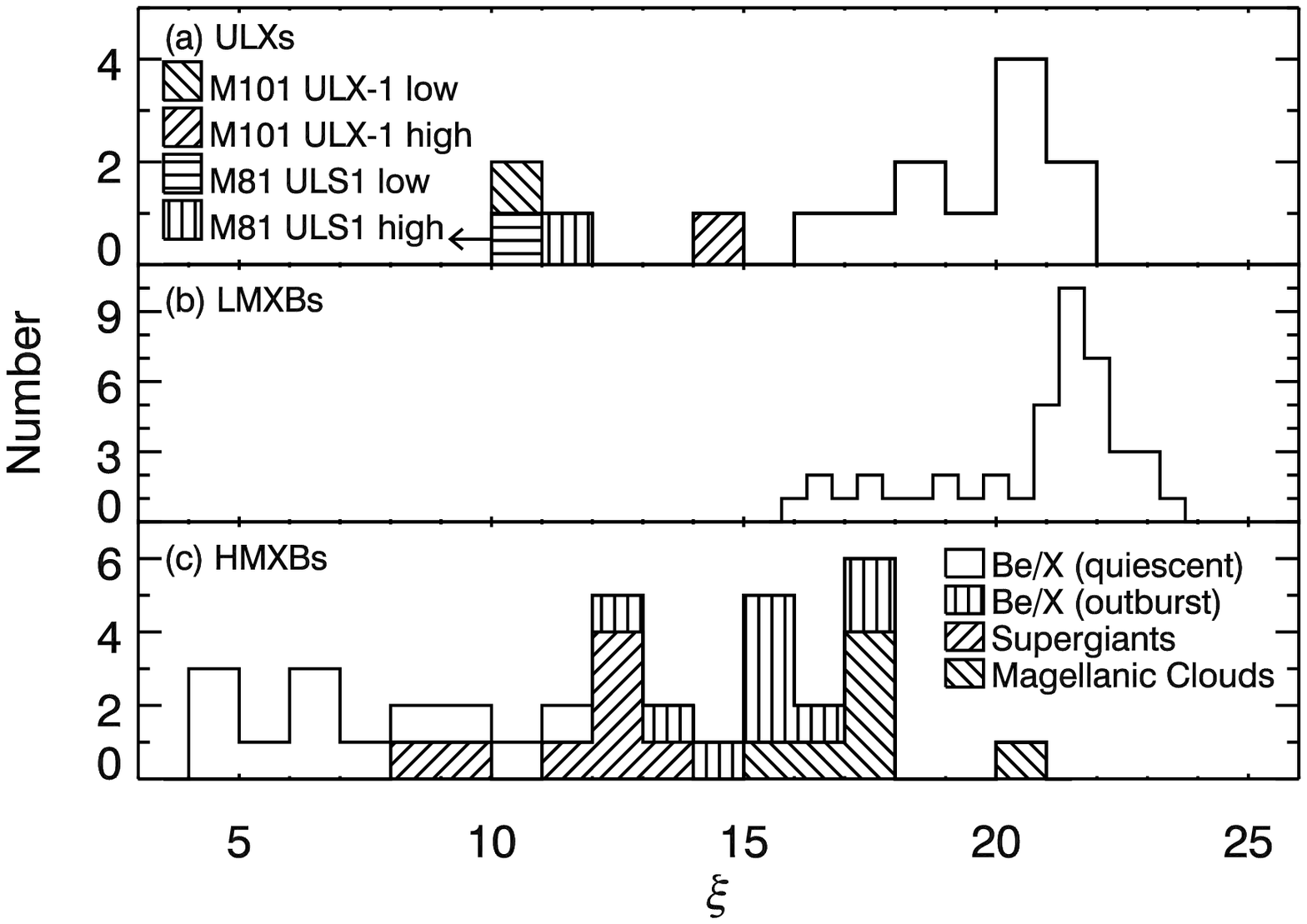}
\caption{Distributions of the X-ray to optical flux ratio, defined as $\log(f_{\rm X}/f_V)$ or $\xi$, for ULXs in the sample, LMXBs, and HMXBs. M81 ULS1 is not detected in 2-10 keV in the low state, and thus the left arrow indicates its upper limit. \label{fig:xor}}
\end{figure}
%%%%%%%%%%%%%%%%%%%%%%%%%%%%%%%%%%%%%%%%%%%%%%%%%%%%%%%%%%%%%%%%%%%%%%%%%%

The X-ray to optical flux ratio, defined following \citet{mac82} as $\log(f_{\rm X}/f_V) = \log f_{\rm X} + m_V/2.5 + 5.37$, where $f_{\rm X}$ is the 0.3-3.5 keV observed flux in $\rm ergs~cm^{-2}~s^{-1}$ and $m_V$ is the visual magnitude, is calculated for each source.  This value can be used to distinguish active galactic nuclei (AGN), BL Lac objects, clusters of galaxies, normal galaxies, normal stars, and X-ray binaries \citep{sto91}. Another form of the X-ray to optical flux ratio as $\xi = B_0 + 2.5 \log F_{\rm X}$ has been proposed to distinguish between low mass X-ray binaries (LMXBs) and high mass X-ray binaries (HMXBs), where $B_0$ is the dereddened $B$ magnitude and $F_{\rm X}$ is the 2-10 keV observed X-ray flux in $\mu$Jy \citep{van95}.

Due to the highly variable nature of these sources, simultaneous X-ray and optical observations may be needed to calculate the flux ratio. Holmberg IX X-1 and NGC 1313 X-2 are the only two sources in our sample that simultaneous X-ray and optical observations are available. For the Chandra observation of Holmberg IX X-1 on 2004 February 7, the pileup effect is manageable; its X-ray flux is derived by fitting the spectrum with a power-law model subject to absorption and pileup. For the Chandra observation of NGC 1313 X-2 on 2004 Feb 22, the source is placed on the chip with a large offset, such that the pileup is negligible; an absorbed power-law model is used to fit the spectrum.

M101 ULX-1 and M81 ULS1 are the two most variable X-ray sources in our sample. The typical X-ray luminosity of M101 ULX-1 is about $2 \times 10^{37}$~erg~s$^{-1}$ in the low state in 0.3-7 keV \citep{kon04}, or a bolometric luminosity of $3 \times 10^{39}$~erg~s$^{-1}$ in the high state \citep{muk05}. For M81 ULS1, the 0.3-8 keV flux is $9.3\times 10^{-15}$~erg~s$^{-1}$~cm$^{-2}$ in the low state and $5.3\times 10^{-13}$~erg~s$^{-1}$~cm$^{-2}$ in the high state \citep{liu08b}. The optical flux does not appear to vary as much as in the X-ray band. We thus calculate two sets of $\log(f_{\rm X}/f_V)$ and $\xi$ using the high and low state X-ray fluxes, respectively. As M81 ULS1 and M101 ULX-1 are very soft and radiate most of their fluxes below 2 keV, the 2-10 keV flux calculated from models in the literature may be inaccurate as model parameters were mainly obtained by fitting the spectrum in the low energy band. Therefore, we measured Chandra count rates in the 2-7 keV band and converted these to fluxes in the 2-10 keV band.  This conversion was done using the power-law index (but not normalization) of the best-fit models from the literature and response matrices calculated at the source position. M81 ULS1 is not detected in the 2-10 keV band in the low state, we thus adopted an upper limit on the flux calculated from the measured upper limit on the count rate.

For other sources, their X-ray flux variability is usually less than a factor of 10 or even 5, which does not affect the X-ray to optical flux ratio significantly. We thus adopted a typical model for each source from the literature\footnote{
Holmberg II X-1, IC 342 X-1, and NGC 5204 X-1 \citep{fen09};
M81 X-6 \citep{miz01};
M83 IXO 82 \citep{sto06};
NGC 1313 X-2 \citep{fen07};
NGC 2403 X-1 \citep{fen05};
NGC 4559 X-7 \citep{sto06};
NGC 5408 X-1 \citep{kaa09b};
NGC 6946 ULX-1 \citep{kaj09}.}
and calculated the flux given the above energy bands. The X-ray fluxes and the X-ray to optical flux ratios are listed in Table~\ref{tab:sum}

The distribution of $\log(f_{\rm X}/f_V)$ and $\xi$ of ULXs in the sample are plotted in Figure~\ref{fig:xor}. The $\xi$ distributions of LMXBs and HMXBs from  \citet{van95} are displayed for comparison.  We note that the 3 of the 13 LMXBs and 3 of the 7 HMXBs in the comparison sample are black hole candidates. For HMXBs, different types are separately indicated as is in the reference. $\xi$ of most ULXs in the sample is consistent with that of LMXBs, and is statistically higher than the distribution of HMXBs.  The X-ray to optical ratios of M101 ULX-1 and M81 ULS1 at the low state are significantly lower than those of other ULXs. At the low state, their $\log(f_{\rm X}/f_V)$ is around zero, consistent with that of AGN, and their $\xi$ is lower than that of LMXBs but similar to HMXBs.

\section{Discussion}
\label{sec:dis}

We have analyzed the largest sample to date of HST observations of ULXs with a unique compact optical counterpart including 3 new identifications and 10 counterparts identified in the literature. For each source, we measured the broad band optical spectral energy distribution (SED), its variation, and the X-ray to optical flux ratio. In the following, we discuss constraints on the nature of ULXs and their optical emission derived from our results.

\subsection{Origin of the optical emission}

X-ray variability rules out that ULXs are clusters of galaxies, normal galaxies, or normal stars.  The $\log(f_{\rm X}/f_V)$ distribution of ULX in our sample, peaked around 3, suggests that most of them are X-ray binaries instead of AGN or BL Lac objects, whose $\log(f_{\rm X}/f_V)$ ranges from $-1$ to 1.2 or 0.3 to 1.7, respectively \citep{sto91}.  

Optical light from an X-ray binary can arise from the companion star, direct emission from the accretion disk, reprocessing of X-rays from the inner disk in the outer disk, or some combination of the above.  If the optical light is dominated by the companion star, then the optical colors in all observations should be consistent with a single spectral type and the optical variability should be low. The only variable stars with colors and magnitudes similar to those of most ULX counterparts are $\beta$-Cepheid and PV Telescopii stars; such star have variabilities of 0.1 mag or less \citep{sta05,jef08}; even heated companion stars, which would vary in an ellipsoidal pattern, are ruled out due to the irregular variation of most ULXs. If the optical light is dominated by direct emission from a multi-color disk blackbody (MCD), then the optical SED should have a $\nu^{1/3}$ form consistent with extrapolation of the X-ray spectrum and there should be significant optical variability.  If the optical light is dominated by reprocessing of X-ray in the outer accretion disk, then the optical SED should have a roughly power-law form with an index between about $-1$ to $2$ \citep{gie09} and, again, there should be significant optical variability, correlated with the X-ray variability.  Also, the X-ray to optical flux ratio should be similar to those observed from LMXBs, in which the optical light is usually dominated by reprocessing of X-ray in the outer accretion disk when in active states.  If the companion and disk produce comparable optical fluxes, then the variability would be reduced relative to that expected from a disk alone and the SED would be a combination of a roughly power-law and a blackbody form.  The observed $\xi$, variability, stellar spectrum fitting, and spectral index are used as evaluation criteria for different models in Table~\ref{tab:model}. These properties favor the irradiated disk model for most sources.

The distributions of $\xi$, $(B-V)_0$, and $(U-B)_0$ of ULXs in the sample appear to be similar to LMXBs, see Figure~\ref{fig:color_histogram} and \ref{fig:xor}. This does not mean that ULXs actually contain low mass companions, but suggests that the optical spectra of ULXs are dominated by emission from the disk instead of the companion star. The fact that the whole optical SED of most ULXs cannot be matched with a single spectral type, see Table~\ref{tab:sp}, is further evidence that the optical emission is not dominated by the companion stars.

Some sources show significant optical variability: M101 ULX-1, M81 ULS1, NGC 1313 X-2, and NGC 5204 X-1.  Notably, both sources for which there are more than 10 observations, M101 ULX-1 and NGC 1313 X-2, show significant variability.  The non-detection of optical variability in the other sources may be due to the limited number of observations available.  Further, these two sources also show color variations that cannot be explained as solely due to changes in extinction, there must be intrinsic changes in the optical spectra.  The detected optical variability is 0.2 mag or larger.  The only variable stars with colors and magnitudes similar to those of the ULX counterparts are $\beta$-Cepheid and PV Telescopii stars.  However, such star have variabilities of 0.1 mag or less \citep{sta05,jef08}. Thus, the variability is inconsistent with a scenario in which the companion star dominates the optical light.  The variability is consistent with the assertion that the optical light from ULXs is dominated by emission from the disk.

The optical spectral index $\alpha$, see Figure~\ref{fig:alpha}, for most of the ULXs in our sample lies in the range $\alpha = -1$ to $2$.  Irradiated disks can produce optical spectra with such slopes \citep{gie09}.  A spectrum with $\alpha = 2$ corresponds to the Rayleigh-Jeans tail of thermal emission and would be produced if all of the optically emitting regions have temperatures above $\sim 2 \times 10^{4}$~K corresponding to the shortest wavelength of the observed optical photons.  Spectral indexes between $-1$ and $2$ suggest that there are optically emitting regions of the disk with temperatures down to $\sim 6000$~K.  Variations in the optical brightness and spectral slope of the reprocessed emission could be produced by variations in the X-ray flux from of the central object, the disk radius, or the fraction of flux intercepted by the outer disk.

A B0 Ib supergiant companion has been suggested for NGC 5204 X-1 based on UV spectroscopy  \citep{liu04}. We re-examined the data but found that the Si~{\sc iii}~$1299$\AA\ line that was used for the classification lies close to the strong geocoronal O~{\sc i}~$1304$\AA\ line, and is thus suspicious. A background spectrum also shows a similar absorption feature. Plus, characteristic absorption lines like C~{\sc iv} and Si~{\sc iv} seen in typical B0 Ib spectra do not appear in the spectrum of NGC 5204 X-1. In the current study, the source displays remarkable variation in color or spectral index, see source \#11 in Figure~\ref{fig:alpha}, that could not be produced if the optical emission is dominated by light from the companion star.  Also, the X-ray to optical flux ratio is consistent with that for a LMXB.  We therefore argue that the optical/UV emission of NGC 5204 X-1 is dominated by the accretion disk and not by a supergiant companion star.

\subsection{Size of the optically emitting region}

If the optical light is due to thermal emission, then the size of the emitting region can be estimated from the flux density measured at a particular frequency using Planck's law if the temperature and distance to the source are known.  Assuming a simple blackbody model for the optical emission, the measurements for Holmberg IX X-1 suggest an emitting region size of $\sim 3 \times 10^{12}$~cm for a distance of 3.6~Mpc and a blackbody temperature of $T = 2\times 10^{4}$~K.  The size estimate decreases by a factor of $\sim5$ if the assumed temperature is increased by a factor of 10 to $T = 2\times 10^{5}$~K.  The size depends more sensitively on temperature for lower temperatures; it increases by a factor of 2 if the temperature is decreased to $T = 1\times 10^{4}$~K and a factor of 9 for $T = 5000$~K.  Temperatures lower than 5000~K appear to be excluded by the optical spectral shape for most of the sources.

It is possible to check if the size of the emitting region is consistent with a picture in which the optical emission is due to reprocessing.  Assuming a fraction $\eta$ of the X-ray luminosity $L_X$ is reprocessed, then $\eta = \sigma T^4 A/L_{X}$ where $T$ and $A$ are the temperature and area, respectively, of the reprocessing region and $\sigma$ is the Stefan-Boltzmann constant.  Inserting $T = 2 \times 10^4$~K, $A = 1 \times 10^{24}$~cm$^2$, and $L_{\rm X} = 1 \times 10^{40}$~\ergs, we find $\eta \sim 0.001$.  This is similar to values found for the Galactic black hole X-ray binary XTE J1817$-$330 when in soft spectral states similar to those usually observed from ULXs \citep{gie09}.  Thus, again, the properties of the optical emission appear to be consistent with reprocessing.

The Milky Way black hole binary with the longest orbital period, GRS~1915+105, has a separation of $8\times 10^{12}$~cm and an outer disk radius of $5\times 10^{12}$~cm, comparable to the values discussed here. The 60 day X-ray periodicity seen from M82 X-1 \citep{kaa07} is longer than that of GRS~1915+105, suggesting a larger orbital separation. The binary separation of M82 X-1 is estimated to be a few times $10^{13}$~cm \citep{fen10}.  In general, large accretion disks, of order $10^{12}$~cm, would require long orbital periods.  Following \citet{fra02}, the disk outer radius should be smaller than the circularization radius, $R_c$, for matter flowing through the Lagrangian point towards the black hole.  The orbital period is related to $R_c$ as

\begin{displaymath}
({P \over {\rm day}}) = ({{R_c} \over {\rm 2.9\times 10^{11} \, cm}})^{3/2}
                                          f^{-3/2} 
                                          ( {{M_{X}} \over M_{\odot}} )^{-1/2}
\end{displaymath}

\noindent where $M_X$ is the mass of the accretor and $f = (1+q)^{4/3} [0.5-0.227 \log q ]^4$ is a function of the mass ratio $q = M_C/M_X$, and $M_C$ is the companion mass.  Requiring $R > 10^{12}$~cm, for $M_X = 10 M_{\odot}$ and $q=0.1$, we find that $P >$~11~days.  This would be a $1 M_{\odot}$ companion.  The period is longer if the companion is heavier, e.g. $P > $26~days for
q=0.5.  Shorter periods would be possible for more massive black holes.

\subsection{Comparison with binary synthesis models}

The $M_V$ vs.\ $(B-V)_0$ diagram can be compared directly with that obtained from binary population synthesis, in which they assume the optical spectrum comprised of emission from both a steady thin disk and the companion \citep{mad08}. Based on their models, the most likely color of ULXs is about $(B-V)_0 = -0.3 \sim -0.2$ independent of the mass of the accretor. However, ULXs with an intermediate mass black hole are brighter than those with a super-Eddington stellar mass black hole of the same luminosity, with the most probable $M_V \approx -6$ for the former and $M_V \approx -4$ for the latter. Our results are consistent with their model that most ULXs have a $(B-V)_0$ color about $-0.3$  to $-0.2$. The concentration of most sources at $M_V \approx -6$ would suggest that most ULXs in our sample contain intermediate mass black holes.  However, we note that this result is model dependent.  Other binary evolution models with different assumptions, i.e.\ that some mass is lost from the system rather than all mass being accreted \citep{pat10}, could lead to different conclusions.

\subsection{NGC 2403 X-1 and M83 IXO 82: Intrinsic disk emission?}

NGC 2403 X-1 has an optical spectral index consistent with that expected, $\alpha = 1/3$, for the intrinsic emission of a standard multicolor disk. Assuming $\alpha = 1/3$, its optical spectrum can be written as $F_\nu = 6 \times 10^{-35} \nu^{1/3}$~\fnu. A face-on standard disk at 3.2~Mpc with an inner temperature $kT_{\rm in} = 1$~keV and a bolometric luminosity $L_{\rm bol} = 3 \times 10^{40}$~\ergs, or with $kT_{\rm in} = 0.1$~keV and $L_{\rm bol} = 10^{39}$~\ergs, can produce an optical tail consistent with the observed spectrum.  The measured X-ray luminosity for the source ranges from 3 to $9 \times 10^{39}$~\ergs \citep{fen09} which is in the required range. Thus, the optical emission for this source may be dominated by intrinsic emission from an accretion disk.  However, simultaneous X-ray and optical/UV observations are needed to test this hypothesis.

The optical spectrum of M83 IXO 82 could be consistent with the $\alpha = 1/3$ expected for intrinsic emission of a standard multicolor disk if our assumed reddening is incorrect.  We calculated the reddening assuming only Galactic absorption of $E(B-V) = 0.066$.  Additional intrinsic absorption of $E(B-V) = 0.111$ would make $\alpha$ consistent with $1/3$.  This source shows an absolute visual magnitude, $M_V \sim -3$, and an X-ray luminosity, $L_X \sim 2\times 10^{39}$~\ergs \citep{fen05}, similar to that of NGC 2403 X-1.  Thus, this source is also a good candidate to have optical emission dominated by intrinsic emission from an accretion disk.

\begin{figure}
\includegraphics[width=\columnwidth]{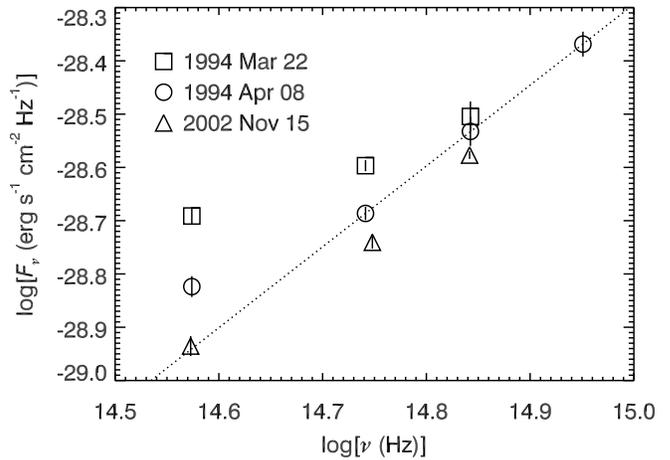}
\caption{Variable optical emission from M101 ULX-1. The dotted line indicates a power-law model fitted to the short wavelength points with $\alpha = 1.52 \pm 0.12$ for the data from 1994 Apr 08.
\label{fig:m101ulx}}
\end{figure}

\subsection{M101 ULX-1 and M81 ULS1}

M101 ULX-1 and M81 ULS1 are exceptional cases owing to their strong X-ray variability and deviation from the main population in the X-ray to optical flux ratio distributions.  Especially when in the low state, their X-ray to optical flux ratios are consistent with that of AGN.  However, detection of variability on time scales of $10^{3}$~s from both sources \citep{muk05,liu08b}, more rapid than the X-ray variability seen from AGN, makes an AGN identification unlikely. The X-ray to optical flux ratio of the two sources is consistent with that of HMXBs.  Many Galactic HXMBs, especially with a neutron star accretor, are transient sources with huge variability due to uneven wind-fed accretion in an elliptical orbit. The high luminosity and moderate absorption column density of these ULXs suggest that they are Roche-lobe overflow system, where circular orbits are often presumed. However, the large long-term variability of these two systems may suggest eccentric orbits.

The optical spectral properties of the two sources are different.  The SED of M101 ULX-1 at 3 epochs is shown in Fig.~\ref{fig:m101ulx}.  There is variability at every waveband where there are multiple observations and the variations are stronger at long wavelengths.  The variability could be explained in a disk irradiation as due to changes in the outer disk radius.  If the blue wing of the spectrum is due to companion light, then a very hot star would be required.  We note that \citet{liu09} fitted the same data using the sum of a Wolf-Rayet star, an irradiated disk described by an $\alpha = -1$ spectrum, and an additional hot component ascribed to the X-ray heated side of the companion.   We note that more complex irradiated disk models \citep{gie09} can produce the extra hot component needed by \citet{liu09} in the inner regions of the disk suffice without need of an additional physical surface.

In contrast, the optical spectrum of M81 ULS1 is significantly shallower.  At least for one observation, the spectrum could be consistent with $\alpha = 1/3$ as expected for intrinsic emission of a standard multicolor disk if there is additional reddening beyond the Galactic value assumed in our analysis. We note that \citet{liu08a} modeled the blue part of the optical spectrum as due to direct disk emission.  However, this modeling was based on simultaneous fitting of HST observations obtained over a period of 4 years.  Given the highly variable nature of the source, it is important to better constrain the spectral behavior with simultaneous observations extending into the UV.

\begin{figure}
\includegraphics[width=\columnwidth]{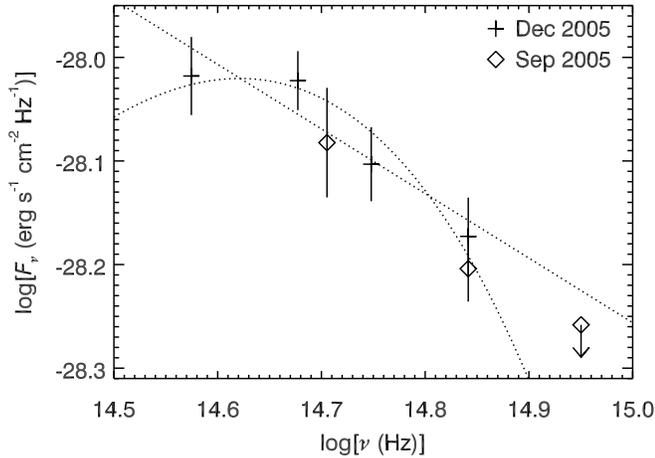}
\caption{Optical spectra of IC 342 X-1 from two sets of quasi-simultaneous observations. The lines are the power-law and blackbody model fitted to the Dec 2005 data. The arrow indicates the 3$\sigma$ upper limit of the flux at F330W where the source is not detected.
\label{fig:ic342}}
\end{figure}

\subsection{IC 342 X-1: A single blackbody?}

The spectrum from IC 342 X-1 on December 2005 can be equally well fitted by a power-law spectrum with $\alpha = -0.62 \pm 0.18$, or a blackbody spectrum with a temperature of $7100 \pm 400$~K, see Figure~\ref{fig:ic342}. However, the spectrum from September 2005 shows a cutoff in the F330W band, inconsistent with the power-law. Fitting to the combined data favors the blackbody model.

Such blackbody emission could arise from a companion star.  There is no evidence of optical variability and a spectral classification of F5 Ib or Iab fits all the data available at both observation epochs.  If the reddening is less than the nebular value we have assumed, then the star would be a later spectral type, i.e.\ F8-G0 Ib \citep{fen08}.  Such blackbody emission could also arise from irradiation in a very large disk.  For IC 342 X-1, the blackbody fitting suggests a surface area of $6 \times 10^{26}$~cm$^2$, or a reflector radius of $10^{13}$~cm, which is about an order of magnitude larger than the value inferred for the majority of ULXs.  Observations at multiple epochs to test for optical variability could distinguish between these two possibilities.

%This is reminiscent of the optically-thin Synchrotron radiation from jets. For IC 342 X-1, two sets of broadband spectra are available from two epochs. Interestingly, IC 342 X-1 happens to have a radio counterpart detected at 5~GHz with a flux of 90~$\mu$Jy \citep[][Phil please check it]{cse10}.  A two point power-law index between the optical and radio is $\alpha_{\rm OR} = -0.2$, much flatter than that in the optical. Therefore, we argue that the optical emission is not due to optically-thin Synchrotron radiation from relativistic jets.

\begin{figure}
\includegraphics[width=\columnwidth]{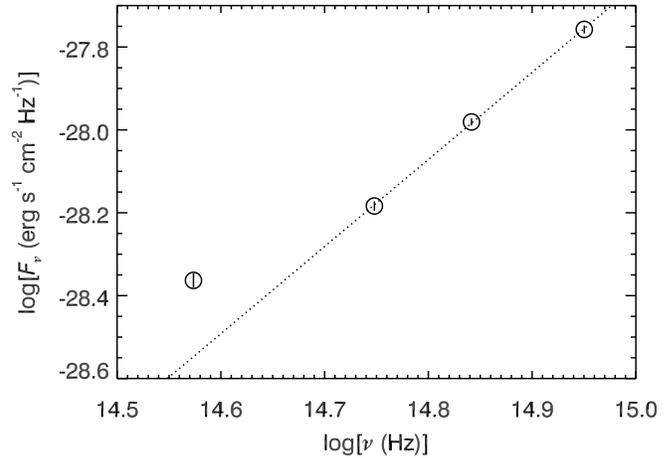}
\caption{The spectrum of Holmberg IX X-1 shows a red excesses in the F814W band. The dotted line indicates a power-law model fitted to the short wavelength points, with $\alpha=2.10 \pm 0.23$.
\label{fig:holmix}}
\end{figure}

\subsection{Holmberg IX X-1: Red excess}

The optical spectrum of Holmberg IX X-1 is well fitted at short wavelengths by a single power-law with a spectral index $\alpha=2.10 \pm 0.23$.  This may represent the Rayleigh-Jeans tail of reprocessed disk emission with temperatures above about $2\times 10^{4}$~K.  At the red end, the spectrum shows an excess at the F814W band above the extrapolation of the power-law spectrum at short wavelengths (Figure~\ref{fig:holmix}).  This indicates some source of emission at cooler temperatures and could be due either to the companion star or some structure in the disk at large radii.  New observations with multiple bands at several epochs could distinguish between these two possibilities.

\subsection{NGC 5204 X-1: not associated with nearby nebulae?}

NGC 5204 X-1 appears on the sky near an optical nebulae \citep{pak02} with a measured nebular extinction of $E(B-V) = 0.4$ \citep{abo07,rus10}. The Galactic extinction along the line of sight of the source is $E(B-V) = 0.013$, which is adopted for reddening correction in this paper.  The source spectrum is found to vary and the power-law index $\alpha$ is around 2.0, which is the largest value theoretically allowed.  Adopting the nebular extinction would lead to an unphysical steep spectral slope, with $\alpha > 2.8$. The X-ray absorption column density of the source is about $(0.6-1.0) \times 10^{21}$~cm$^{-2}$ \citep{fen09}, which translates to an extinction of $E(B-V) = 0.11-0.22$ \citep{pre95} allowing for the range of dust-to-gas ratios discussed in \citet{car89} and \citet{pre95}. This is significantly lower than the nebular value. Therefore, we suggest that the X-ray source may be not associated with the nebulae.

\section{Conclusions}
\label{sec:con}

\begin{itemize}
\setlength{\itemsep}{0pt}
\setlength{\parskip}{0pt} 
\setlength{\leftskip}{1em}
  \item The dominant component of optical emission in most ULXs is produced via X-ray irradiation of the outer accretion disk, similar to the case of LMXBs.  
  \item As a corollary, one cannot use broad-band optical colors to infer the companion spectral type.
  \item Large optical emitting regions are required, suggesting relatively large orbital separations and long orbital periods.
  \item The optical spectrum and X-ray luminosity of NGC 2403 X-1 are consistent with both arising from intrinsic emission from a standard accretion disk.  M83 IXO 82 is a somewhat weaker candidate for intrinsic disk emission extending into the optical.
  \item M101 ULX-1 and M81 ULS1 are unique in high X-ray variability and low X-ray to optical flux ratios. These systems may have eccentric orbits.
  \item IC 342 X-1 has a spectrum quite different from that of most ULXs and the optical emission may be dominated by an F5 or somewhat later supergiant companion star.
  \item NGC 5204 X-1 seems not to be physically related to the nearby optical nebulae due to inconsistent extinction. 
\end{itemize}

\acknowledgments We thank the anonymous referee for insightful comments. HF acknowledges funding support from the National Natural Science Foundation of China under grant No.\ 10903004 and 10978001, the 973 Program of China under grant 2009CB824800, the Foundation for the Author of National Excellent Doctoral Dissertation of China under grant 200935, the Tsinghua University Initiative Scientific Research Program, and the Program for New Century Excellent Talents in University.  PK and FG acknowledge partial support from NASA grant NNX08AJ26G.

{\it Facility:} \facility{CXO}, \facility{HST}

\clearpage

%%%%%%%%%%%%%%%%%%%%%%%%%%%%%%%%%%%%%%%%%%%%%%%%%%%%%%%%%%%%%%%%%%%%%%%%%%
\begin{deluxetable}{lllllllll}
\tablewidth{0pc} 
\setlength{\tabcolsep}{3pt}
\tabletypesize{\scriptsize}
\tablecaption{Absolute visual magnitudes, colors, and spectral classifications of the ULXs in the sample
\label{tab:sp}}
\tablehead{
\colhead{ULX} & \colhead{date} & \colhead{$M_V$} & \colhead{$\ub$} & \colhead{$\bv$} & \colhead{$V-I$} & \colhead{$V-R$} & \colhead{Sp.} & \colhead{$\chi^2/{\rm dof}$}}
\startdata
Holmberg II X-1 & 2006-12-30 & $-6.00 $ & \nodata & \nodata   & $-0.02$      & \nodata       & B6-B7 Ib-Iab  &B0 I, 112.4/8; B0.5 I, 70.6/8;\\
                & 2007-10-03 & $-5.93 $ & \nodata & $-0.24$   & \nodata      & \nodata       & B0-B1 Ib      &B5 I, 73.5/8  \\
                & 2007-10-05 & $-5.94 $ & \nodata & $-0.18$   & \nodata      & \nodata       & B1-B2 Ib      &  \\
                & 2007-10-09 & $-5.94 $ & \nodata & $-0.27$   & \nodata      & \nodata       & O9-B0 Ib      &  \\

Holmberg IX X-1  &2004-02-07& $-5.88$ & $-1.33$   &  $-0.42$   &  $ -0.04^{\star}$    & \nodata   & O5V, O6III &O3 V, 232.3/4; O5 V, 253.3/4  \\

IC 342 X-1  &2005-09-02& $-5.94$ & $>-0.77$ &  $0.37 $   & \nodata      & \nodata       & F5 Ib-Iab  & F5 I, 10.7/5    \\
            &2005-12-18& $-5.95$ & \nodata  &  $0.30 $   &  $0.60$      &  $0.38 $      & F5 Ib-Iab  &     \\

M81 ULS1  &2006-03-21 & $-6.26$ & \nodata         &  $0.11$  & \nodata       & \nodata   & A5-A7 Iab  &F2 I, 424.1/6; F5 I, 418.3/6;  \\
          &2006-03-27 & $-5.99$ & \nodata         &  $0.22$  & \nodata       & \nodata   & F2 Ib-Iab  &F8 I, 786.3/6; G0 I, 1505.1/6 \\

M81 X-6  &1995-01-31  &$-3.91$ & $-1.51^{\star}$ & $0.06$    &  $-0.05$         &  $-0.06$ &  B5-A3 II   &B2 II, 78.6/14; B5 II, 61.6/14    \\
         &2001-06-04  &$-4.01$ & \nodata         & $-0.13$   &  $0.07^{\star}$  & \nodata  &  B7-B8 II   &    \\

M83 IXO 82       &2006-02-25& $-3.11$ & $-0.96^{\star}$&  $0.26 $  & \nodata  & \nodata   & A2-F2 II, B1 V  &F0 II, 16.2/2; F2 II, 11.5/2;   \\
                 &          &         &                &           &          &           &                 &B1 V, 21.7/2   \\

M101 ULX-1 & 1994-03-22 & $-6.39 $ & \nodata &  $-0.11$   &  $0.20^{\star}$   & \nodata&  B5 Iab, A3 Iab          &O5 I, 412.8/9; O8 III, 342.3/9; \\
           & 1994-04-08 & $-6.15 $ & $-1.20$ &  $-0.28$   &  $0.10^{\star}$   & \nodata&  O5 III, O8 II           &B5 I, 320.1/9  \\
           & 2002-11-15 & $-6.00 $ & \nodata &  $-0.31$   &  $-0.01^{\star}$  & \nodata&  O5 III, O8-O9 II, B7 Ib &  \\

NGC 1313 X-2 & 2003-11-22 & $-4.61 $ & $-1.14$ &  $-0.26$   &  $-0.09^{\star}$     & \nodata&   O9V, B0-B1 III   &O9 V, 76.4/3; B0 III, 62.7/3;   \\
             &            &          &         &            &                      &        &                    &B1 III, 153.1/3   \\

NGC 2403 X-1     &2005-10-17& $-2.90$ & $-1.01 $ &  $0.07^{\star}$& \nodata& \nodata   & B0-B2 V &B0 V, 33.6/2; B1 V, 28.5/2; \\
                 &          &         &          &                &        &           &         &B2 V, 33.3/2    \\

NGC 4559 X-7  &2001-05-25& $-6.98 $  & \nodata &  $-0.17$   &  $-0.11$   & \nodata   & B2-B5 Ia   &B5 I, 26.0/6     \\
              &2005-03-08& $-7.10 $  & \nodata &  $-0.12$   &  $-0.06$   & \nodata   & B3-B5 Ia   &    \\

NGC 5204 X-1  &2001-05-28 & $-5.77 $ & \nodata   & \nodata   & $-0.52$       & \nodata   & O5V         &O9 III, 33.8/9; B0 I, 25.7/9; \\
              &2008-08-08 & $-5.57 $ & \nodata   &  $-0.35$  & \nodata       & \nodata   & O5V, O8III  &O5 V, 707.5/9; B2 II, 1060.8/9 \\
              &2008-08-10 & $-5.62 $ & \nodata   &  $-0.23$  & \nodata       & \nodata   & B2 II-Ib    & \\
              &2008-08-13 & $-5.69 $ & \nodata   &  $-0.20$  & \nodata       & \nodata   & B3 II-Ib    & \\

NGC 5408 X-1 & 2009-04-04 & $-6.27$ & $-1.11$ &  $-0.28$   &  $-0.19^{\star}$    & \nodata       & O9 Ib     &O8 I, 29.8/5; O9.5 I, 37.1/5   \\

NGC 6946 ULX-1  &1996-01-27& $-7.26$        & \nodata &  $-0.07$         & \nodata      & \nodata   & B6-B7 Ia    &O5 I, 8.3/5; O8 I, 7.3/5;    \\
                &2001-06-08& $-7.27 $       & \nodata &  $-0.42^{\star}$ &  $ -0.21$    & \nodata   & B2 Ia       &B0 I, 4.3/5; B5 I, 6.0/5
\enddata
%\tablenotetext{$\star$}{Color that does not agree with the best-fit spectral type.}
\tablecomments{$^{\star}$Color that does not agree with the best-fit spectral type.}
\end{deluxetable}
%%%%%%%%%%%%%%%%%%%%%%%%%%%%%%%%%%%%%%%%%%%%%%%%%%%%%%%%%%%%%%%%%%%%%%%%%%

%%%%%%%%%%%%%%%%%%%%%%%%%%%%%%%%%%%%%%%%%%%%%%%%%%%%%%%%%%%%%%%%%%%%%%%%%%
\begin{deluxetable}{lllrllllllc}
\tablewidth{0pc} 
\setlength{\tabcolsep}{2pt}
\tabletypesize{\scriptsize}
\tablecaption{Optical and X-ray Properties of the ULXs in the sample
\label{tab:sum}}
\tablehead{
\colhead{ULX} &\colhead{date} &\colhead{$M_V$} &\colhead{$\alpha$} &\colhead{$\chi^2/{\rm dof}$} &\colhead{$f_{\rm X}$} &$F_{\rm X}$ &\colhead{$\log(f_{\rm X}/f_V)$} &\colhead{$\xi$} &\colhead{$V_{\max} - V_{\min}$} &\colhead{\# obs}}

\startdata
Holmberg II X-1$^*$ &2006-12-30&$-6.00$   &$1.10\pm0.09$& \nodata    &4.08E$-$12 &1.09E$-$01 &2.7    &18.9   &$0.07\pm0.03$ & 4\\
                &2007-10-03&$-5.93$   &$1.41\pm0.22$& \nodata    & \nodata        & \nodata        & \nodata      & \nodata      & \nodata & \nodata \\
                &2007-10-05&$-5.94$   &$1.17\pm0.21$& \nodata    & \nodata        & \nodata        & \nodata      & \nodata      & \nodata & \nodata \\
                &2007-10-09&$-5.94$   &$1.52\pm0.19$& \nodata    & \nodata        & \nodata        & \nodata      & \nodata      & \nodata & \nodata \\

Holmberg IX X-1 &2004-02-07&$-5.88$   &$1.79\pm0.19$&44.9/2    &4.92E$-$12 &3.62E$-$01   &3.2    &20.4       &$0.09\pm0.04$ & 2\\

IC 342 X-1  &2005-12-18&$-5.95$   &$-0.62\pm0.18$&2.0/2    &1.79E$-$12  &1.78E$-$01 &3.3    &20.1       &$0.01\pm0.13$ & 2\\

M81 ULS1$^*$    &2006-03-21&$-6.26$   &$0.02\pm0.13$  & \nodata   &5.43E$-$15$^l$   &$<3.51$E$-$05$^l$ &$-0.1$$^l$   &$<10.7^l$       &$0.27\pm0.06$ & 2\\
            &2006-03-27&$-5.99$   &$-0.39\pm0.13$ & \nodata   &2.03E$-$13$^h$   &5.21E$-$05$^h$ &1.4$^h$    &11.1$^h$       & \nodata & \nodata\\

M81 X-6  &1995-01-31&$-3.91$       &$1.19\pm0.18$&3.8/3    &1.55E$-$12 &1.02E$-$01 &3.2    &21.3        &$0.18\pm0.09$ & 5\\
			&2001-06-04&$-4.01$       &$0.94\pm0.14$&0.1/1    & \nodata    & \nodata        & \nodata          & \nodata       & \nodata  & \nodata\\

M83 IXO 82  &2006-02-25&$-3.11$   &$-0.43\pm0.35$&0.4/1    &3.18E$-$13 &8.88E$-$03 &3.0  &20.4       & \nodata & \nodata\\

M101 ULX-1&1994-03-22&$-6.39$       &$0.62\pm0.10$&1.0/1    &2.64E$-$15$^l$  &1.64E$-$05$^l$  &$0.2^l$ &10.9$^l$       &$0.67\pm0.03$ & 27 \\
			 &1994-04-08&$-6.15$       &$1.20\pm0.08$&10.8/2   &3.39E$-$13$^h$   &3.37E$-$04$^h$  &2.3$^h$ &14.2$^h$    & \nodata & \nodata \\
          &2002-11-15&$-6.00$       &$1.41\pm0.07$&7.4/1    & \nodata          & \nodata         & \nodata       & \nodata          & \nodata & \nodata \\

NGC 1313 X-2    &2003-11-22&$-4.61$   &$1.37\pm0.14$&3.5/2   &1.72E$-$12         &1.17E$-$01 &3.1  &20.5       &$0.20\pm0.04$ & 22\\

NGC 2403 X-1    &2005-10-17&$-2.90$   &$0.34\pm0.20$&0.8/1    &8.33E$-$13 &3.81E$-$02 &3.2  &21.1       & \nodata & \nodata \\

NGC 4559 X-7  &2001-05-25&$-6.98$   &$1.25\pm0.07$&1.2/1        &5.89E$-$13    &2.24E$-$02 &2.3  &18.7   &$0.11\pm0.04$ & 3\\
              &2005-03-08&$-7.10$   &$1.06\pm0.10$&0.5/2        & \nodata           & \nodata        & \nodata    & \nodata      & \nodata  & \nodata \\

NGC 5204 X-1$^*$&2001-05-28&$-5.77$   &$2.42\pm0.39$& \nodata    &1.58E$-$12  &4.95E$-$02 &2.6      &19.2 &$0.12\pm0.03$ & 4\\
            &2002-10-28& \nodata  &$0.92\pm0.11$& \nodata    & \nodata         & \nodata        & \nodata        & \nodata    & \nodata & \nodata \\
	    &2008-08-08&$-5.57$   &$1.87\pm0.27$& \nodata    & \nodata         & \nodata        & \nodata        & \nodata    & \nodata & \nodata \\
            &2008-08-10&$-5.62$   &$1.38\pm0.46$& \nodata    & \nodata         & \nodata        & \nodata        & \nodata    & \nodata & \nodata \\
            &2008-08-13&$-5.69$   &$1.26\pm0.28$& \nodata    & \nodata         & \nodata        & \nodata        & \nodata    & \nodata & \nodata \\

NGC 5408 X-1  &2009-04-04&$-6.27$       &$1.41\pm0.14$&3.6/2    &3.32E$-$12   &2.33E$-$02 &2.8  &17.8       &$0.03\pm0.05$ & 2\\

NGC 6946 ULX-1  &2001-06-08&$-7.27$   &$1.88^{+0.48}_{-0.47}$&0.5/1    &4.77E$-$13    &2.09E$-$02 &2.2  &16.8       &$0.01\pm0.17$ & 2
\enddata
\tablecomments{$^*$Sources that have quasi-simultaneous observations available only at two bands; $^l$X-ray low state; $^h$X-ray high state; $\alpha$ is the power-law index of the optical spectrum; $\chi^2/{\rm dof}$ is derived from fitting with a power-law model; $f_{\rm X}$ is the observed 0.3-3.5 keV flux in \ergcms; $F_{\rm X}$ is the 2-10 keV observed X-ray flux in $\mu$Jy; $V_{\max} - V_{\min}$ indicates the largest variation in the $V$ band; \# obs is the number of observations for which a V magnitude was measured.}
\end{deluxetable}
%%%%%%%%%%%%%%%%%%%%%%%%%%%%%%%%%%%%%%%%%%%%%%%%%%%%%%%%%%%%%%%%%%%%%%%%%%

\begin{deluxetable}{llccclccc}
\tablewidth{\columnwidth}
\tabletypesize{\scriptsize}
\tablewidth{0pc}
\tablecolumns{9}
\tablecaption{Evaluation of different models
\label{tab:model}}
\tablehead{
\colhead{ULX} &  \multicolumn{4}{c}{Properties}                                &  & \multicolumn{3}{c}{Models}   \\
\cline{2-5} \cline{7-9}
\colhead{}    & \colhead{$\xi$}    & \colhead{Variability}&  \colhead{Stellar spectrum} & \colhead{Spectral index} &  & \colhead{Star} & \colhead{Irradiated disk}  & \colhead{Direct disk} }
\startdata
Holmberg II X-1                & L & M       &  B    & R  &   & $-$   & $+$  & $-$     \\
Holmberg IX X-1                & L & M       &  B    & R  &   & $-$   & $+$  & $-$     \\
IC 342 X-1                     & L & N       &  G    & R  &   & $++$  & $+$  & $-$     \\
M81 ULS1                       & H & V       &  B    & R  &   & $-$   & $+$  & $-$     \\
M81 X-6                        & L & M       &  B    & R  &   & $-$   & $+$  & $-$     \\
M83  IXO 82                    & L & \nodata &  B    & R  &   & $-$   & $+$  & $-$     \\
M101 ULX-1                     & H & V       &  B    & R  &   & $-$   & $+$  & $-$     \\
NGC 1313 X-2                   & L & V       &  B    & R  &   & $-$   & $++$ & $-$     \\
NGC 2403 X-1                   & L & \nodata &  B    & D  &   & $-$   & $0$  & $+$     \\
NGC 4559 X-7                   & L & M       &  B    & R  &   & $-$   & $+$  & $-$     \\
NGC 5204 X-1                   & L & V       &  B    & R  &   & $-$   & $++$ & $-$     \\
NGC 5408 X-1                   & H & N       &  B    & R  &   & $-$   & $0$  & $-$     \\
NGC 6946 ULX-1                 & H & N       &  G    & R  &   & $++$  & $0$  & $-$     \\

\enddata
\tablecomments{$\xi$: these sources whose $\xi$ are larger than 18 are marked as L, otherwise, are marked as H; Variability: V, M and N are used for these sources whose $V_{\max} - V_{\min} \geqq 4~\rm{error}$,  $4~\rm{error} > V_{\max} - V_{\min} \geqq 2~\rm{error} $ and $V_{\max} - V_{\min}<~\rm{error}$, respectively; Stellar spectrum: using $\chi^2/{\rm dof}$ in Table~\ref{tab:sp} to calculate the probability P that a random variable $\chi^2$ is less than or equal to observed $\chi^2$. The stellar spectrum fitting is bad (B) if $\rm P > 99\%$, otherwise,  is good (G); Spectral index: D is used if $\alpha$ is consistent with 1/3 within errors, otherwise, R is used. Star: $++$ if stellar spectrum fitting is good and variability is N, $-$ if stellar spectrum fitting is bad or variability is V; Irradiated disk: $++$ if $\xi=\rm{L}$ \& spectral index = R \& variability = V, $+$ if $\xi=\rm{L}$ \& spectral index = R \& variability $\neq$ V or $\xi=\rm{H}$ \& spectral index = R \& variability = V,  0 if $\xi=\rm{L}$ \& spectral index $\neq$ R \& variability $\neq$ V or $\xi=\rm{H}$ \& spectral index = R \& variability $\neq$ V; Direct disk: $+$ if spectral index = D, $-$ if spectral index = R. The rate of how strongly the model is recommended should be based on the decreasing from $++$ to $-$.}
\end{deluxetable}

%%%%%%%%%%%%%%%%%%%%%%%%%%%%%%%%%%%%%%%%%%%%%%%%%%%%%%%%%%%%%%%%%%%%%%%%%%

%\appendix

\LongTables % only for emulateapj

\begin{deluxetable}{clllcll}
\tablewidth{0pc}
\setlength{\tabcolsep}{1pt}
\tabletypesize{\scriptsize}
\tablecaption{HST observations and magnitudes of  ULXs in our sample
\label{tab:mag}}
\tablehead{
\colhead{ULX} & \colhead{date}& \colhead{dataset} & \colhead{instrument and filter} & \colhead{exposure (s)} & \colhead{$m_{\rm Filter}$} & \colhead{$m_V$}
}
\startdata
Holmberg II X-1	&2006-01-28& j9dr03010           &ACS/WFC/F814W    &600&$21.450\pm0.029$   & \\
                &2006-12-30& j9cmz7010           &ACS/WFC/F555W    &4660&$21.403\pm0.030$  &  $21.418\pm0.030$\\
                &2006-12-30& j9cmz7020           &ACS/WFC/F814W    &4660&$21.451\pm0.019$  & \\
                &2007-10-03& u9zs240$\rm [1|2]m$ &WFPC2/PC1/F450W  &500/500        &$21.287\pm0.018$  & \\
                &2007-10-03& u9zs240$\rm [3|4]m$ &WFPC2/PC1/F555W  &500/500        &$21.471\pm0.012$  &  $21.489\pm0.012$\\
                &2007-10-05& u9zs250$\rm [1|2]m$ &WFPC2/PC1/F450W  &500/500        &$21.328\pm0.015$  & \\
                &2007-10-05& u9zs250$\rm [3|4]m$ &WFPC2/PC1/F555W  &500/500        &$21.465\pm0.012$  &  $21.478\pm0.012$\\
                &2007-10-09& u9zs260$\rm [1|2]m$ &WFPC2/PC1/F450W  &500/500        &$21.259\pm0.014$  & \\
                &2007-10-09& u9zs260$\rm [3|4]m$ &WFPC2/PC1/F555W  &500/500        &$21.464\pm0.012$  &  $21.484\pm0.012$\\
\noalign{\smallskip}\hline\noalign{\smallskip}
Holmberg IX X-1 &2004-02-07& j8ol03010     & ACS/WFC/F814W &1160&$21.885\pm0.042$ & \\
                &2004-02-07& j8ol03040     & ACS/WFC/F435W &2520&$21.461\pm0.021$ & \\
                &2004-02-07& j8ola3010     & ACS/HRC/F330W &2760&$19.616\pm0.024$ & \\
                &2004-02-07& j8ol03030     & ACS/WFC/F555W &1160&$21.868\pm0.028$ & $21.901\pm0.028$\\
&2004-03-25& j8ol07010     & ACS/WFC/F555W &2400&$21.960\pm0.021$ & $21.992\pm0.021$\\
\noalign{\smallskip}\hline\noalign{\smallskip}
IC 342 X-1  &2005-09-02&j9h8b3010           & ACS/HRC/F330W    &2900&$>20.868$       & \\
            &2005-09-02&j9h813010           & ACS/WFC/F435W    &1248&$22.019\pm0.132$&   \\
&2005-09-02&j9h813020           & ACS/WFC/F606W    &1248&$21.521\pm0.079$&  $21.651\pm0.100$ \\
            &2005-12-16&j9jn$\rm [0|a|b|c]1010$ & ACS/WFC/F625W    &920/920/920/920        &$21.293\pm0.071$& \\
            &2005-12-18&j9jn02020           & ACS/WFC/F814W    &1080&$21.022\pm0.094$&\\
            &2005-12-18&j9jn02040               & ACS/WFC/F435W    &1800&$21.942\pm0.094$&   \\
            &2005-12-18&j9jn02010           & ACS/WFC/F555W    &1080&$21.667\pm0.089$&  $21.640\pm0.089$ \\
\noalign{\smallskip}\hline\noalign{\smallskip}
M81 ULS1 &2003-09-18& j8mx18akq      &ACS/WFC/F814W     &120&$20.965\pm0.050$ & \\
         &2004-09-15& j90la7010      &ACS/WFC/F814W     &1650&$21.019\pm0.051$ & \\
         &2005-12-09& u9el200$\rm[3|4]m$ &WFPC2/WF4/F300W       &500/500       &$20.369\pm0.075$ & \\
         &2006-03-21& j9el15020      &ACS/WFC/F606W     &1200&$21.460\pm0.034$ & $21.541\pm0.044$\\
&2006-03-21& j9el15010      &ACS/WFC/F435W     &1200&$21.648\pm0.020$ & \\
         &2006-03-27& j9el28a9q      &ACS/WFC/F435W     &465&$22.026\pm0.024$ & \\
         &2006-03-27& j9el28aaq      &ACS/WFC/F606W     &470&$21.699\pm0.032$ & $21.811\pm0.041$\\
\noalign{\smallskip}\hline\noalign{\smallskip}
M81 X-6  &1995-01-31& u2mh010$\rm [b|c]t$           & WFPC2/WF4/F555W & 300/600         &   $23.880\pm0.060$    &    $23.894\pm0.060$\\
         &1995-01-31& u2mh010$\rm [d|e]t$           & WFPC2/WF4/F675W & 300/600         &   $23.958\pm0.107$    &   \\
         &1995-01-31& u2mh010$\rm [5|6|7]$t         & WFPC2/WF4/F336W & 400/160/600     &   $21.939\pm0.142$    &   \\
&1995-01-31& $\rm u2mh010[f|g]t$           & WFPC2/WF4/F814W & 300/600         &   $23.955\pm0.109$    &   \\
         &1995-01-31& u2mh010$\rm [8|9|a]$t         & WFPC2/WF4/F439W & 400/400/400     &   $23.982\pm0.166$    &   \\
         &2001-06-04& u6eh010$\rm [5r|6m|7r|8r]$    & WFPC2/PC1/F450W & 500/500/500/500 &   $23.674\pm0.043$    &   \\
         &2001-06-04& u6eh010$\rm [1|2|3|4]r$       & WFPC2/PC1/F555W & 500/500/500/500 &   $23.782\pm0.060$    &    $23.791\pm0.060$\\
         &2001-06-04& $\rm u6eh010[9|a|b|c]r$       & WFPC2/PC1/F814W & 500/500/500/500 &   $23.737\pm0.077$    &   \\
         &2004-09-15& j90l08010                 & ACS/WFC/F814W   &1650&   $23.560\pm0.099$    &     \\
         &2004-09-25& u8ZO010$\rm [1|3]m$           & WFPC2/WF3/F675W & 600/600         &   $23.618\pm0.074$    &    \\
         &2004-09-25& u8ZO010$\rm [2|4]m$           & WFPC2/WF3/F555W & 500/600         &   $23.879\pm0.076$    &    $23.872\pm0.076$\\
         &2006-03-22& j9el18010                 & ACS/WFC/F435W   &1200&   $23.722\pm0.029$    &    \\
         &2006-03-22& j9el18020                 & ACS/WFC/F606W   &1200&   $23.664\pm0.054$    &     $23.717\pm0.068$\\
         &2006-03-27& j9ela8aeq                 & ACS/WFC/F606W   &450&   $23.789\pm0.053$    &     $23.829\pm0.068$\\
         &2006-03-27& j9ela8acq                 & ACS/WFC/F435W   &400&   $23.781\pm0.043$    &    \\
\noalign{\smallskip}\hline\noalign{\smallskip}
M83 IXO82 &2006-02-25& j9h811020      &ACS/WFC/F606W    &1000&$25.132\pm0.107$ &\\
&2006-02-25& j9h811010      &ACS/WFC/F435W    &1000&$25.512\pm0.080$ &\\
          &2006-02-25& j9h8a1010      &ACS/HRC/F330W    &2568&$24.156\pm0.263$ &\\
\noalign{\smallskip}\hline\noalign{\smallskip}
M101 ULX1 &1994-03-22&u2780101t            & WFPC2/WF4/F555W &1200&$22.895\pm0.026$  &  $22.898\pm0.026$ \\
          &1994-03-22&u2780102t            & WFPC2/WF4/F814W &1000&$22.712\pm0.041$  &\\
          &1994-03-23&u2780103t            & WFPC2/WF4/F439W &1000&$22.813\pm0.070$  &\\
          &1994-03-30&u2780201t            & WFPC2/WF4/F555W &1200&$23.022\pm0.032$  &  $23.040\pm0.032$\\
          &1994-04-08&u2780302t            & WFPC2/WF4/F814W &1200&$23.044\pm0.049$  &\\
          &1994-04-08&u2780301t            & WFPC2/WF4/F555W &1200&$23.120\pm0.031$  &  $23.133\pm0.031$\\
          &1994-04-08&u2780303t            & WFPC2/WF4/F439W &1200&$22.884\pm0.068$  &\\
          &1994-04-08&u278030$\rm [4|5]t$  & WFPC2/WF4/F336W &    1200/1200   &$21.143\pm0.058$  &\\
          &1994-04-11&u278040$\rm [1p|2t]$ & WFPC2/WF4/F555W &    1200/120    &$23.009\pm0.059$  &  $23.027\pm0.059$\\
          &1994-04-16&u2780601t            & WFPC2/WF4/F555W &1200&$23.108\pm0.031$  &  $23.126\pm0.031$\\
          &1994-04-20&u2780701t            & WFPC2/WF4/F555W &1200&$23.160\pm0.032$  &  $23.178\pm0.032$\\
          &1994-04-24&u2780801t            & WFPC2/WF4/F555W &1200&$23.192\pm0.040$  &  $23.210\pm0.040$\\
          &1994-04-28&u2780901t            & WFPC2/WF4/F555W &1200&$23.167\pm0.028$  &  $23.185\pm0.028$\\
          &1994-05-03&u2780A01t            & WFPC2/WF4/F555W &1200&$23.222\pm0.033$  &  $23.240\pm0.033$\\
          &1994-05-05&u2780501t            & WFPC2/WF4/F555W &1200&$23.249\pm0.031$  &  $23.267\pm0.031$\\
          &1994-05-05&u278050$\rm [2|3]t$  & WFPC2/WF4/F814W &    1200/120    &$23.252\pm0.070$  &\\
          &1994-05-10&u2780$\rm [b|c]01t$  & WFPC2/WF4/F555W &    1200/1200   &$23.256\pm0.022$  &  $23.274\pm0.022$\\
          &1994-05-10&u2780b02t            & WFPC2/WF4/F814W &1200&$23.316\pm0.053$  &\\
          &1995-03-22&u2780d01t            & WFPC2/WF4/F555W &1200&$23.237\pm0.034$  &  $23.255\pm0.034$\\
          &1995-04-17&u2ms030$\rm [1|2]t$  & WFPC2/WF4/F555W &    500/500     &$23.208\pm0.039$  &  $23.226\pm0.039$\\
          &2002-11-15&j8d601031            & ACS/WFC/F814W   &720&$23.315\pm0.047$  &\\
          &2002-11-15&j8d601021            & ACS/WFC/F555W   &720&$23.261\pm0.030$  &  $23.283\pm0.030$\\
          &2002-11-15&j8d601011            & ACS/WFC/F435W   &900&$22.951\pm0.018$  &\\
&2006-12-23&j9o401020            & ACS/WFC/F555W   &1330&$23.177\pm0.014$  &  $23.199\pm0.014$\\
          &2006-12-23&j9o401010            & ACS/WFC/F814W   &724&$23.146\pm0.033$  &\\
          &2006-12-24&j9o402020            & ACS/WFC/F555W   &1330&$23.167\pm0.015$  &  $23.190\pm0.015$\\
          &2006-12-24&j9o402010            & ACS/WFC/F814W   &724&$23.116\pm0.029$  &\\
          &2006-12-25&j9o403020            & ACS/WFC/F555W   &1330&$23.205\pm0.015$  &  $23.227\pm0.015$\\
          &2006-12-25&j9o403010            & ACS/WFC/F814W   &724&$23.144\pm0.027$  &\\
          &2006-12-26&j9o404020            & ACS/WFC/F555W   &1330&$23.181\pm0.014$  &  $23.204\pm0.014$\\
          &2006-12-26&j9o404010            & ACS/WFC/F814W   &724&$23.183\pm0.032$  &\\
          &2006-12-28&j9o405020            & ACS/WFC/F555W   &1330&$23.156\pm0.015$  &  $23.178\pm0.015$\\
          &2006-12-28&j9o405010            & ACS/WFC/F814W   &724&$23.227\pm0.030$  &\\
          &2006-12-30&j9o406020            & ACS/WFC/F555W   &1330&$23.201\pm0.016$  &  $23.224\pm0.016$\\
          &2006-12-30&j9o406010            & ACS/WFC/F814W   &724&$23.119\pm0.030$  &\\
          &2007-01-01&j9o407020            & ACS/WFC/F555W   &1330&$23.243\pm0.015$  &  $23.266\pm0.015$\\
          &2007-01-01&j9o407010            & ACS/WFC/F814W   &724&$23.216\pm0.030$  &\\
          &2007-01-04&j9o408020            & ACS/WFC/F555W   &1330&$23.249\pm0.016$  &  $23.272\pm0.016$\\
          &2007-01-04&j9o408010            & ACS/WFC/F814W   &724&$23.221\pm0.030$  &\\
          &2007-01-07&j9o409020            & ACS/WFC/F555W   &1330&$23.311\pm0.017$  &  $23.334\pm0.017$\\
          &2007-01-07&j9o409010            & ACS/WFC/F814W   &724&$23.177\pm0.031$  &\\
          &2007-01-11&j9o410020            & ACS/WFC/F555W   &1330&$23.179\pm0.016$  &  $23.202\pm0.016$\\
          &2007-01-11&j9o410010            & ACS/WFC/F814W   &724&$23.069\pm0.028$  &\\
          &2007-01-17&j9o411020            & ACS/WFC/F555W   &1330&$23.009\pm0.012$  &  $23.031\pm0.012$\\
          &2007-01-17&j9o411010            & ACS/WFC/F814W   &724&$22.928\pm0.025$  &\\
          &2007-01-21&j9o412020            & ACS/WFC/F555W   &1330&$23.081\pm0.014$  &  $23.104\pm0.014$\\
          &2007-01-21&j9o412010            & ACS/WFC/F814W   &724&$23.054\pm0.031$  &\\
          &2008-01-05&ua30640$\rm [3|4]m$  & WFPC2/WF3/F555W &    500/500     &$22.642\pm0.024$  &  $22.660\pm0.024$\\
          &2008-01-05&ua30640$\rm [1|2]m$  & WFPC2/WF3/F814W &350&$22.616\pm0.048$  &\\
\noalign{\smallskip}\hline\noalign{\smallskip}
NGC 1313 X-2    &2003-11-22&   j8ola2010           &ACS/HRC/F330W      &2760&   $21.317\pm0.029$    &       \\
        &2003-11-22&   j8ol02040           &ACS/WFC/F435W      &2520&   $22.955\pm0.012$    &       \\
        &2003-11-22&   j8ol02010           &ACS/WFC/F814W      &1160&   $23.339\pm0.023$    &       \\
        &2003-11-22&   j8ol02030           &ACS/WFC/F555W      &1160&   $23.211\pm0.017$    &       $23.233\pm0.017$\\
        &2004-02-22&   j8ol06010           &ACS/WFC/F555W      &2400&   $23.086\pm0.015$    &       $23.108\pm0.015$\\
        &2008-05-21&   u9zs010$\rm [3|4]m$ &WFPC2/PC1/F555W    &   400/700 &   $23.057\pm0.028$    &       $23.074\pm0.028$    \\
        &2008-05-21&   u9zs010$\rm [1|2]m$ &WFPC2/PC1/F450W    &   500/500 &   $22.982\pm0.044$    &       \\
        &2008-05-22&   u9zs020$\rm [3|4]m$ &WFPC2/PC1/F555W    &   400/700 &   $23.149\pm0.033$    &       $23.166\pm0.033$    \\
        &2008-05-22&   u9zs020$\rm [1|2]m$ &WFPC2/PC1/F450W    &   500/500 &   $22.881\pm0.054$    &       \\
        &2008-05-23&   u9zs030$\rm [1|2]m$ &WFPC2/PC1/F450W    &   500/500 &   $22.863\pm0.046$    &       \\
        &2008-05-23&   u9zs030$\rm [3|4]m$ &WFPC2/PC1/F555W    &   400/700 &   $23.115\pm0.034$    &       $23.132\pm0.034$    \\
        &2008-05-24&   u9zs040$\rm [3|4]m$ &WFPC2/PC1/F555W    &   400/700 &   $23.153\pm0.030$    &       $23.170\pm0.030$    \\
        &2008-05-24&   u9zs040$\rm [1|2]m$ &WFPC2/PC1/F450W    &   500/500 &   $23.029\pm0.044$    &       \\
        &2008-05-25&   u9zs050$\rm [3|4]m$ &WFPC2/PC1/F555W    &   400/700 &   $23.212\pm0.031$    &       $23.229\pm0.031$    \\
        &2008-05-25&   u9zs050$\rm [1|2]m$ &WFPC2/PC1/F450W    &   500/500 &   $23.056\pm0.053$    &       \\
        &2008-05-26&   u9zs060$\rm [1|2]m$ &WFPC2/PC1/F450W    &   500/500 &   $23.030\pm0.049$    &       \\
        &2008-05-26&   u9zs060$\rm [3|4]m$ &WFPC2/PC1/F555W    &   400/700 &   $23.254\pm0.038$    &       $23.271\pm0.038$    \\
        &2008-05-27&   u9zs070$\rm [1|2]m$ &WFPC2/PC1/F450W    &   500/500 &   $23.073\pm0.046$    &       \\
        &2008-05-27&   u9zs070$\rm [3|4]m$ &WFPC2/PC1/F555W    &   400/700 &   $23.188\pm0.031$    &       $23.205\pm0.031$    \\
        &2008-05-28&   u9zs080$\rm [3|4]m$ &WFPC2/PC1/F555W    &   400/700 &   $23.188\pm0.032$    &       $23.205\pm0.032$    \\
        &2008-05-28&   u9zs080$\rm [1|2]m$ &WFPC2/PC1/F450W    &   500/500 &   $22.994\pm0.047$    &       \\
        &2008-05-29&   u9zs090$\rm [3|4]m$ &WFPC2/PC1/F555W    &   400/700 &   $23.109\pm0.029$    &       $23.126\pm0.029$    \\
        &2008-05-29&   u9zs090$\rm [1|2]m$ &WFPC2/PC1/F450W    &   500/500 &   $22.836\pm0.041$    &       \\
        &2008-05-30&   u9zs100$\rm [1|2]m$ &WFPC2/PC1/F450W    &   500/500 &   $22.923\pm0.048$    &       \\
        &2008-05-30&   u9zs100$\rm [3|4]m$ &WFPC2/PC1/F555W    &   400/700 &   $23.075\pm0.029$    &       $23.092\pm0.029$    \\
        &2008-05-31&   u9zs110$\rm [1|2]m$ &WFPC2/PC1/F450W    &   500/500 &   $23.014\pm0.050$    &       \\
        &2008-05-31&   u9zs110$\rm [3|4]m$ &WFPC2/PC1/F555W    &   400/700 &   $23.215\pm0.032$    &       $23.232\pm0.032$    \\
        &2008-06-01&   u9zs120$\rm [3|4]m$ &WFPC2/PC1/F555W    &   400/700 &   $23.135\pm0.033$    &       $23.152\pm0.033$    \\
        &2008-06-01&   u9zs120$\rm [1|2]m$ &WFPC2/PC1/F450W    &   500/500 &   $23.049\pm0.046$    &       \\
        &2008-06-02&   u9zs130$\rm [3|4]m$ &WFPC2/PC1/F555W    &   400/700 &   $23.251\pm0.033$    &       $23.268\pm0.033$    \\
        &2008-06-02&   u9zs130$\rm [1|2]m$ &WFPC2/PC1/F450W    &   500/500 &   $23.085\pm0.048$    &       \\
        &2008-06-03&   u9zs140$\rm [3|4]m$ &WFPC2/PC1/F555W    &   400/700 &   $23.132\pm0.038$    &       $23.149\pm0.038$    \\
        &2008-06-03&   u9zs140$\rm [1|2]m$ &WFPC2/PC1/F450W    &   500/500 &   $22.980\pm0.043$    &       \\
        &2008-06-04&   u9zs150$\rm [3|4]m$ &WFPC2/PC1/F555W    &   400/700 &   $23.261\pm0.034$    &       $23.278\pm0.034$    \\
        &2008-06-04&   u9zs150$\rm [1|2]m$ &WFPC2/PC1/F450W    &   500/500 &   $22.977\pm0.051$    &       \\
        &2008-06-05&   u9zs160$\rm [3|4]m$ &WFPC2/PC1/F555W    &   400/700 &   $23.182\pm0.031$    &       $23.199\pm0.031$    \\
        &2008-06-05&   u9zs160$\rm [1|2]m$ &WFPC2/PC1/F450W    &   500/500 &   $22.908\pm0.052$    &       \\
        &2008-06-06&   u9zs170$\rm [1|2]m$ &WFPC2/PC1/F450W    &   500/500 &   $23.073\pm0.051$    &       \\
        &2008-06-06&   u9zs170$\rm [3|4]m$ &WFPC2/PC1/F555W    &   400/700 &   $23.196\pm0.031$    &       $23.213\pm0.031$    \\
        &2008-06-07&   u9zs180$\rm [3|4]m$ &WFPC2/PC1/F555W    &   400/700 &   $23.153\pm0.033$    &       $23.170\pm0.033$    \\
&2008-06-07&   u9zs180$\rm [1|2]m$ &WFPC2/PC1/F450W    &   500/500 &   $23.149\pm0.053$    &       \\
        &2008-06-08&   u9zs190$\rm [3|4]m$ &WFPC2/PC1/F555W    &   400/700 &   $23.247\pm0.034$    &       $23.264\pm0.034$    \\
        &2008-06-08&   u9zs190$\rm [1|2]m$ &WFPC2/PC1/F450W    &   500/500 &   $23.030\pm0.056$    &       \\
        &2008-06-09&   u9zs200$\rm [3|4]m$ &WFPC2/PC1/F555W    &   400/700 &   $23.253\pm0.037$    &       $23.270\pm0.037$    \\
        &2008-06-09&   u9zs200$\rm [1|2]m$ &WFPC2/PC1/F450W    &   500/500 &   $22.953\pm0.043$    &       \\
\noalign{\smallskip}\hline\noalign{\smallskip}
NGC 2403 X-1  &2005-10-17& j9h803020     &ACS/WFC/F606W    &1248&$24.564\pm0.064$ &\\
&2005-10-17& j9h803010 &ACS/WFC/F435W    &1248&$24.687\pm0.040$ &\\
              &2005-10-17& j9h8a3010     &ACS/HRC/F330W    &2912&$23.243\pm0.171$ &\\
\noalign{\smallskip}\hline\noalign{\smallskip}
NGC 4559 X-7 &2001-05-25&u6eh050$\rm [9|a|b|c]r$ & WFPC2/PC1/F814W &500/500/500/500 &$23.139\pm0.038$ & \\
             &2001-05-25&u6eh050$\rm [5|6|7|8]r$ & WFPC2/PC1/F450W &500/500/500/500 &$22.878\pm0.023$ & \\
             &2001-05-25&u6eh050$\rm [1|2|3|4]r$ & WFPC2/PC1/F555W &500/500/500/500 &$23.002\pm0.020$ & $23.016\pm0.020$ \\
             &2005-03-08&j92w01040               & ACS/HRC/F814W   &2400&$22.981\pm0.083$& \\
             &2005-03-08&j92w01020               & ACS/HRC/F555W   &2400&$22.888\pm0.033$ & $22.902\pm0.033$ \\
&2005-03-08&j92w01010               & ACS/HRC/F435W   &2400&$22.744\pm0.019$ & \\
             &2005-03-09&j92w02030           & ACS/WFC/F606W   &1828&$22.929\pm0.102$ & $22.938\pm0.108$ \\
\noalign{\smallskip}\hline\noalign{\smallskip}
NGC 5204 X-1 &2001-05-28& u6723901r            &WFPC2/WF2/F606W    &600&$22.492\pm0.044$&  $22.393\pm0.089$\\
             &2001-05-28& u6723902r            &WFPC2/WF2/F814W    &600&$22.925\pm0.103$& \\
             &2002-10-28& j8dq01010            &ACS/HRC/F435W      &2600&$22.271\pm0.067$& \\
             &2002-10-29& j8dq01020            &ACS/HRC/F220W      &2720&$19.851\pm0.024$& \\
             &2008-08-08& u9zs210$\rm [1|2]m$  &WFPC2/PC1/F450W    &500/500        &$22.301\pm0.022$& \\
&2008-08-08& u9zs210$\rm [3|4]m$  &WFPC2/PC1/F555W    &500/500        &$22.572\pm0.021$&  $22.601\pm0.021$\\
             &2008-08-10& u9zs220$\rm [1|2]m$  &WFPC2/PC1/F450W    &500/500        &$22.350\pm0.023$& \\
             &2008-08-10& u9zs220$\rm [3|4]m$  &WFPC2/PC1/F555W    &500/500        &$22.528\pm0.019$&  $22.545\pm0.019$\\
             &2008-08-13& u9zs230$\rm [1|2]m$  &WFPC2/PC1/F450W    &500/500        &$22.307\pm0.023$& \\
             &2008-08-13& u9zs230$\rm [3|4]m$  &WFPC2/PC1/F555W    &500/500        &$22.463\pm0.018$&  $22.478\pm0.018$\\
\noalign{\smallskip}\hline\noalign{\smallskip}
NGC 5408 X-1    &2000-07-04& u6724101r                  & WFPC2/WF3/F606W   &600&$22.175\pm0.020$ &  $22.161\pm0.041$\\
                &2000-07-04& u6724102r                  & WFPC2/WF3/F814W   &600&$22.266\pm0.040$ & \\
                &2009-04-04& ubah290$\rm [1|2|3]m$      & WFPC2/WF4/F439W   & 600/600/600       &$21.879\pm0.028$ &\\
                &2009-04-04& ubah290$\rm [4|5|6|7]m$    & WFPC2/WF4/F336W   & 1100/1100/1100/1100         &$20.216\pm0.016$ &\\
                &2009-04-04& ubah290$\rm [bm|dm|fn]$    & WFPC2/WF4/F555W   & 180/180/160       &$22.114\pm0.022$ &  $22.132\pm0.022$\\
                &2009-04-04& ubah290$\rm [cm|en|gn]$    & WFPC2/WF4/F814W   & 180/180/160       &$22.339\pm0.045$ & \\
\noalign{\smallskip}\hline\noalign{\smallskip}
NGC 6946 ULX-1  &1996-01-27& u33d0105t                  & WFPC2/PC1/F555W   &400&$21.268\pm0.115$ &  $21.275\pm0.115$\\
                &1996-01-27& u33d010$\rm [6|7]t$        & WFPC2/PC1/F439W   & 400/400               &$21.239\pm0.184$ &\\
&2001-06-08& u6eh020$\rm [5|6|7|8]r$    & WFPC2/PC1/F450W   & 500/500/500/500       &$20.905\pm0.130$ &\\
                &2001-06-08& u6eh020$\rm [1|2|3|4]r$    & WFPC2/PC1/F555W   & 500/500/500/500       &$21.238\pm0.122$ &  $21.267\pm0.122$\\
                &2001-06-08& u6eh020$\rm [9|a|b|c]r$    & WFPC2/PC1/F814W   & 500/500/500/500       &$21.494\pm0.103$ & \\
                &2004-07-29& j8mxd8sxq                  & ACS/WFC/F814W     &120&$21.450\pm0.072$ &
\enddata

\tablecomments{$m_{\rm Filter}$ is the magnitude quoted in the filter bandpass. $m_V$ is the magnitude in Johnson $V$, translated from F555W or F606W. All magnitudes are in VEGA zero point corrected for extinction.}
\end{deluxetable}

\end{document}